\documentclass[aps,prd,longbibliography,preprintnumbers,nofootinbib,
superscriptaddress,amsmath,floatfix,10pt]{revtex4}

\usepackage{amssymb}  
\usepackage{amsmath}
\usepackage{hyperref}
\usepackage{graphicx
}
\usepackage{exscale}
\usepackage{bbold}
\usepackage{textcomp}
\usepackage{enumerate}
\usepackage [english]{babel}
\usepackage[utf8]{inputenc}
\usepackage [autostyle, english = american]{csquotes}
\MakeOuterQuote{"}

\usepackage{color}

\usepackage[normalem]{ulem}  

\newcommand{\tr}{\text{Tr}}

\newcommand{\non}{\nonumber\\}

\newcommand{\be}{\begin{equation}}
\newcommand{\ee}{\end{equation}}
\newcommand{\bea}{\begin{eqnarray}}
\newcommand{\eea}{\end{eqnarray}}
\newcommand{\ba}[1]{\begin{array}{#1}}
\newcommand{\ea}{\end{array}}

\begin{document}

\title{
Chiral crossover vs chiral density wave in dense nuclear matter
}

\author{Savvas Pitsinigkos}
\email{S.Pitsinigkos@soton.ac.uk}
\affiliation{Mathematical Sciences and STAG Research Centre, University of Southampton, Southampton SO17 1BJ, United Kingdom}

\author{Andreas Schmitt}
\email{a.schmitt@soton.ac.uk}
\affiliation{Mathematical Sciences and STAG Research Centre, University of Southampton, Southampton SO17 1BJ, United Kingdom}

\date{22 April 2024}


\begin{abstract} 
We employ a model based on nucleonic and mesonic degrees of freedom to discuss the competition between isotropic and anisotropic phases in cold and dense matter. Assuming isotropy, the model exhibits a chiral phase transition which is of second order in the chiral limit and becomes a crossover in the case of a realistic pion mass. This observation crucially depends on the presence of the nucleonic vacuum contribution. Allowing for an anisotropic phase in the form of a chiral density wave can disrupt the smooth crossover. We identify the regions in the parameter space of the model where a chiral density wave is energetically preferred. A  high-density re-appearance of the chiral density wave with unphysical behavior, as seen in previous studies, is avoided by a suitable renormalization scheme. 
A nonzero pion mass tends to disfavor the anisotropic phase 
compared to the chiral limit and we find that, within our model, the chiral density wave is only realized for baryon densities of at least about 6 times nuclear saturation density. 

\end{abstract}

\maketitle


\section{Introduction}
\label{sec:intro}

\subsection{Background and motivation}

Thermodynamic phases that break rotational and/or translational invariance are ubiquitous in condensed-matter systems and are expected to play an important role in the phase diagram of Quantum Chromodynamics (QCD). Cold and dense matter governed by QCD can be found inside neutron stars and thus the properties of anisotropic or crystalline phases are important for the understanding of astrophysical data. Neutron stars rotate and contain strong magnetic fields, effects that tend to stabilize anisotropic structures on a microscopic level. But, even without any external fields, cold and dense matter is prone to developing spatial structures, typically because a condensation mechanism becomes ``stressed''. A non-uniform state can then be stabilized as a result of competing effects, finding a balance between keeping the kinetic energy cost small while sustaining a gain from condensation energy. In cold and dense quark matter, a mismatch in Fermi momenta due to the nonzero strange quark mass puts a stress
on the uniform quark/quark pairing, resulting in anisotropic or crystalline Cooper pair condensates \cite{Rajagopal:2005dg,Alford:2007xm,Anglani:2013gfu}.  Here we will be concerned with the possibility of an anisotropic {\it chiral} condensate. In this case, the baryon chemical potential itself imposes a stress on the condensation mechanism because chiral condensation is based on quark/anti-quark pairing. Throughout the paper we will ignore the possibility of Cooper pairing for simplicity and consider systems without magnetic field or rotation.

Since the anisotropic  state is an intermediate phase between chirally broken and (approximately) chirally restored phases, we expect a spatially varying chiral condensate in the vicinity of the chiral transition\footnote{It is conceivable that an anisotropic or inhomogeneous chiral condensate persists up to asymptotically large densities -- then in the form of quark/quark-hole pairing. However, in QCD this requires a large number of colors \cite{Deryagin:1992rw,Shuster:1999tn}.}. Chiral (and deconfinement) transitions are strong-coupling phenomena and cannot be described with perturbative methods. Moreover, in the region of cold and dense matter, even brute-force methods on the lattice are currently inapplicable. Therefore, for now, this regime of QCD is inaccessible from first principles.  The discussion of the chiral transition in cold and dense matter is thus mostly limited to phenomenological models, including the study of inhomogeneous phases in its vicinity. The vast majority of these studies have been performed in models based on quark degrees of freedom, such as the Nambu--Jona-Lasinio (NJL) or quark-meson model \cite{Nakano:2004cd,Nickel:2009wj,Frolov:2010wn,Carignano:2011gr,Carignano:2014jla,Buballa:2015awa,Adhikari:2016eef,Adhikari:2017ydi,Andersen:2018osr,Ferrer:2019zfp,Carignano:2019ivp,Lakaschus:2020caq,Buballa:2020xaa}. These models are, at best, suitable for the high-density side of the chiral transition. However, the relevant degrees of freedom on the low-density side, where chiral symmetry is spontaneously broken, are nucleons. Therefore, these models only yield a toy version of the chirally broken phase of cold and dense QCD. Ideally,  a model should account for both quark and nuclear matter. This is a very challenging task, even on the less rigorous level of phenomenological models.  Attempts include models where quark and nucleon degrees of freedom are combined in the Lagrangian \cite{Negreiros:2010hk} and models based on the gauge-gravity duality, where both confined and deconfined phases arise naturally but which are different from real-world QCD in other aspects \cite{BitaghsirFadafan:2018uzs,Ishii:2019gta,Kovensky:2020xif}. 

\subsection{Model and main idea}

Here, we employ a nucleon-meson model \cite{Drews:2014spa}, which offers a complementary perspective to NJL and quark-meson models: In this approach, the low-density side does contain the correct degrees of freedom (and we can choose the model parameters to reproduce properties of nuclear matter at saturation). On the other hand, we have to live with a toy version of quark matter. One of the main ideas is that, if combined, the two complementary approaches can give solid predictions for QCD, at least on a qualitative level. Importantly, our model does have a chiral limit, and thus knows about the concept of a chiral phase transition. The reason is that the nucleon mass is generated fully dynamically, in contrast to widely used models for dense nuclear matter that  contain a mass parameter in the Lagrangian, such as the Walecka model and its variants \cite{Walecka:1974qa,Boguta:1977xi,Serot:1984ey,Alford:2020pld}.  The model we employ was already used 
in the context of the chiral transition, for instance to compute the surface tension of the interface between the two phases \cite{Fraga:2018cvr}, and strangeness was  included to account for a somewhat more realistic description of the chirally symmetric phase \cite{Fraga:2022yls}. It was also used to construct mixed phases under neutron star conditions \cite{Schmitt:2020tac}. A mixed phase is another example of a spatially inhomogeneous structure, with  spatial regions, for instance bubbles, of one phase immersed in the background of another phase. In the context of the quark-hadron transition, the possibility of mixed phases is closely related to a first-order transition in the presence of a local charge neutrality constraint and the relaxation of this constraint to global neutrality. In this paper, we restrict ourselves to isospin-symmetric nuclear matter without any neutrality condition, where these mixed phases play no role. 

Instead, we will allow for an anisotropic chiral condensate, which oscillates between scalar and pseudoscalar components along a certain, spontaneously chosen, direction in position space -- this is commonly referred to as a chiral density wave (CDW)\footnote{Other names for the CDW exist in the literature, such as ``axial wave condensation'', ``dual chiral density wave'', or ``chiral spiral''. The analogue in quark/quark pairing (as opposed to  quark/anti-quark pairing in the chiral condensate) is referred to as the single plane wave Larkin-Ovchinnikov-Fulde-Ferrell (LOFF) state. The CDW is also conceptually the same as a superfluid with nonzero superflow in a fixed direction, described by a complex scalar field whose phase varies along this direction (which can be visualized as a spiral) \cite{Schmitt:2014eka}.}. The CDW has been used as an ansatz for the chiral condensate in numerous studies because of its simplicity. In particular, it does not break translational invariance for any observable. More complicated structures have been discussed, for instance spatial variations in the scalar component only \cite{Nickel:2009wj,Buballa:2020xaa}, variants of the chiral density wave \cite{Takeda:2018ldi},  higher-dimensional lattice structures \cite{Abuki:2011pf,Carignano:2012sx}, all reviewed in  Ref.\ \cite{Buballa:2014tba}, and the possibility of a quantum spin liquid \cite{Pisarski:2020dnx}. It is not the purpose of our study to compare these different inhomogeneous phases, hence we need to keep in mind that our CDW phase may itself be unstable with respect to a phase that does break translational invariance.

\subsection{Main novelties}

The chiral density wave in nuclear matter was already analyzed in  Refs.\ \cite{Dautry:1979bk,Takahashi:2001jq,Takahashi:2002md}, employing a model similar to ours, and in Refs.\ \cite{Heinz:2013hza,Takeda:2018ldi}, where an extended linear sigma model was used, describing nucleons in a parity doublet and, in Ref.\ \cite{Heinz:2013hza}, including an additional scalar field. All these works ignore the vacuum contribution of the nucleons (the ``Dirac sea''), and we will argue that this contribution makes an important difference. This difference is important not only if a CDW is included. (In Ref.\ \cite{Haber:2014ula} it was argued that the Dirac sea is crucial in the presence of a magnetic field.)  Even for the isotropic case, the location and nature of the chiral phase transition is corrected significantly by the vacuum contribution, as already pointed out in the model we use here \cite{Brandes:2021pti}. In models with quark degrees of freedom, on the other hand, the Dirac sea {\it was} included together with the CDW, at least in some of the above mentioned works, see for instance Refs.\  \cite{Carignano:2014jla,Andersen:2018osr}. Implementing this contribution for the first time into a study of the CDW in nuclear matter gives us a more realistic picture. Moreover, we carefully discuss the renormalization needed for the Dirac sea and point out that a suitable renormalization procedure avoids artifacts at high density seen previously in NJL and quark-meson approaches \cite{Carignano:2011gr,Carignano:2014jla}. Additionally, we will show how the CDW is affected by a quartic self-coupling of the vector meson \cite{Sugahara:1993wz}, which was not taken into account in Refs.\ \cite{Dautry:1979bk,Takahashi:2001jq,Takahashi:2002md,Heinz:2013hza,Takeda:2018ldi}, but which has recently been explored to account for realistic neutron stars \cite{Dexheimer:2020rlp,Providencia:2023rxc}. We also ask whether a CDW is favored in a system which -- in the absence of anisotropic phases -- shows a smooth chiral crossover. As we shall see, the crossover is an unavoidable consequence of our model if the Dirac sea  is included, and thus we are ``forced'' to work in this scenario, which is a viable possibility in QCD \cite{Baym:2017whm}. And we shall see that an anisotropic chiral condensate can indeed introduce phase transitions in an otherwise smooth crossover. This is not unlike the Bose-Einstein condensation/Bardeen-Cooper-Schrieffer crossover \cite{Schmitt:2014eka}, which can also be disrupted by phase transitions if there is a mismatch in Fermi momenta for the two fermion species that form pairs \cite{Sheehy_2007,Deng:2006ed}. 

\subsection{Structure of the paper}

Our paper is organized as follows. In Sec.~\ref{sec:model} we introduce our model of isospin-symmetric nuclear matter and incorporate the CDW, see Secs.\ \ref{sec:lag} and \ref{sec:ansatz}. Then, in Secs.\ \ref{sec:free} and \ref{sec:stat} we derive the free energy and set up the stationarity equations, including the Dirac sea contribution, which requires renormalization, explained in detail in Appendix \ref{sec:diracsea}. We explain our procedure for fitting the 
model parameters in Sec.\ \ref{sec:fit}. Our main results are presented in Sec.~\ref{sec:results}, starting from the effect of the vacuum terms on the isotropic scenario in Sec.\ \ref{sec:isotropic}. The CDW is studied in Sec.~\ref{sec:res_chiral_limit} for a specific parameter set, before we present a more global view of the parameter space in Sec.\ \ref{sec:phase_structure}. We compare our results to previous approaches in the literature regarding the treatment of the Dirac sea in Sec.\ \ref{sec:compare}, before we give a summary and an outlook in Sec.~\ref{sec:summary}.

\section{Model and ansatz}\label{sec:model}

\subsection{Lagrangian}
\label{sec:lag}
Our model is based on a Lagrangian containing baryonic, mesonic, and interaction terms \cite{Drews:2014spa,Fraga:2018cvr,Schmitt:2020tac,Brandes:2021pti,Fraga:2022yls},
\be
\mathcal{L} = \mathcal{L}_{\rm bar} + \mathcal{L}_{\rm mes} + \mathcal{L}_{\rm int} \, . 
\ee
The baryonic part is 
\begin{equation}
{\cal L}_{\rm bar} = \bar{\psi}(i\gamma^\mu\partial_\mu+
\gamma^0\mu)\psi	\, ,		
\end{equation}
where the nucleon spinor contains neutrons and protons, $\psi=(\psi_n, \psi_p)$, and $\mu$ is the baryon chemical potential. Throughout the paper, we will restrict ourselves to isospin-symmetric nuclear matter, where neutrons and protons are degenerate and in particular have the same chemical potential, $\mu_n=\mu_p\equiv \mu$. The Lagrangian does not include a nucleonic mass parameter, the nucleon mass will be generated dynamically by spontaneous chiral symmetry breaking. The mesonic part is 
\begin{equation} \label{LM}
{\cal L}_{\rm mes} = \,\frac{1}{2}\partial_\mu\sigma\partial^\mu\sigma
 +\frac{1}{4}\tr [\partial_\mu \pi \partial^\mu \pi]
  -\frac{1}{4}\omega_{\mu\nu}\omega^{\mu\nu}+\frac{m_\omega^2}{2}\omega_\mu\omega^\mu
 +\frac{d}{4}\left(\omega_\mu\omega^\mu\right)^2 - {\cal U}(\sigma, \pi)	\, ,	 	
\end{equation}
where $\pi = \pi_a \tau_a$ with the Pauli matrices $\tau_a$ is the pion field, where $\omega_{\mu\nu}\equiv \partial_\mu \omega_\nu - \partial_\nu\omega_\mu$, where $m_\omega = 782\, {\rm MeV}$ is the vector meson mass, and where $d>0$ is the (dimensionless) self-coupling constant of the vector meson. The potential for the sigma and pion fields takes the form 
\begin{equation} \label{Usigpi}
{\cal U}(\sigma, \pi) = \sum_{n=1}^4 \frac{a_n}{n!} 
\frac{(\sigma^2+\pi_a\pi_a-f_\pi^2)^n}{2^n}-\epsilon(\sigma-f_\pi) \, ,
\end{equation}
with parameters $a_1, a_2,a_3, a_4, \epsilon$, and the pion decay constant $f_\pi=93 \, {\rm MeV}$. The potential incorporates a (small) explicit chiral symmetry breaking through the parameter $\epsilon$, which is proportional to the pion mass. For $\epsilon=0$ the Lagrangian is invariant under chiral transformations. Finally, baryons and mesons are assumed to interact via the Yukawa interaction
\begin{equation}
{\cal L}_{\rm int} = -\bar{\psi}\left[g_{\sigma} (\sigma +i\gamma^5 \pi )+ g_{\omega}
\gamma^\mu \omega_\mu  \right]\psi		\, , 
\end{equation}
with coupling constants $g_\sigma$ and $g_\omega$.

\subsection{Ansatz and mean-field approximation}
\label{sec:ansatz}

In the simplest situation, only the fields $\sigma$ and $\omega^0$ develop expectation values. We separate them from the fluctuations, $\sigma \to \phi + \sigma$, $\omega^0\to\omega + \omega^0$, where $\phi \equiv \langle\sigma\rangle$, $\omega\equiv \langle\omega^0\rangle$ are density-dependent (and in general also temperature-dependent) condensates. If we assume isotropy, $\phi$ and $\omega$ are constant in space. In our more general -- anisotropic -- ansatz we keep the vector meson condensate $\omega$ spatially constant and introduce a spatial modulation in the sector of scalar and pseudoscalar condensates in the form of a CDW, 
\begin{equation} \label{CDWansatz}
{\sigma} = \phi \cos(2 \vec{q}\cdot \vec{x}), \quad {\pi}_3 = \phi \sin( 2 \vec{q}\cdot \vec{x}) \, . 
\end{equation}
Here, the wave vector $\vec{q}$ breaks rotational invariance spontaneously, and its modulus $q$ as well as the condensate $\phi$ have to be determined dynamically. We have set any charged pion condensate to zero; its competition and possible coexistence with the CDW is worth exploring in systems with isospin asymmetry \cite{Dautry:1979bk,Andersen:2018osr}.  
It is useful to work with transformed fermionic fields according to
\begin{equation} \label{trafo}
\psi \to e^{-i\gamma^5\tau_3 \vec{q}\cdot\vec{x}} \psi \, .
\end{equation}
Using this transformation and neglecting the mesonic fluctuations, we can write the ``mean-field Lagrangian'' as 
\begin{equation} \label{LMF}
\mathcal{L}_\text{mf} = \bar{\psi}(i\gamma_\mu\partial^\mu + \gamma^0 \mu_*-M+\gamma^5\vec{q}\cdot\vec{\gamma}\,\tau_3)\psi 
+ \frac{m_\omega^2}{2}{\omega}^2+\frac{d}{4}{\omega}^4  - U - \Delta U \, , 
\end{equation}
where we have introduced the effective nucleon mass,
\begin{equation}
M = g_\sigma \phi \, , 
\end{equation}
and the effective chemical potential,
\begin{equation}
\mu_* = \mu-g_\omega {\omega} \, .
\end{equation}
The mesonic vacuum potential is written as a sum of isotropic and 
$q$-dependent contributions,  
\begin{equation}
U\equiv U(\phi) \equiv  \sum_{n=1}^4 \frac{a_n}{n!} \frac{(\phi^2-f_\pi^2)^n}{2^n}-\epsilon(\phi-f_\pi)  \, , \qquad \Delta U\equiv \Delta U(\phi,q) \equiv  2\phi^2q^2 + (1-\delta_{0q})\,\epsilon\phi \, . \label{UDelU}
\end{equation}
The form of the $q$-dependent part deserves a comment. First of all, the term $2\phi^2q^2$  originates from the kinetic term in Eq.\ (\ref{LM}) and corresponds to a kinetic energy cost of creating an axial current from the mesonic sector. Next, we notice that the explicit symmetry breaking term in the potential (\ref{Usigpi}) retains a spatial dependence that cannot be transformed away by the fermionic transformation (\ref{trafo}). This spatial dependence is easy to understand: Even without the presence of nucleons the CDW ansatz (\ref{CDWansatz}) is only a solution to the Euler-Lagrange equations for $\sigma$ and $\pi_3$ in the chiral limit $\epsilon=0$; in that case, the solution traces a circle in the $\sigma$-$\pi_3$ plane. An explicit symmetry breaking term $\epsilon>0$ ``tilts'' the vacuum potential and the solution will no longer be circular; the condensate will ``wobble'' along the spatial direction parallel to $\vec{q}$ rather than smoothly follow a regular spiral\footnote{We have solved the Euler-Lagrange equations for $\sigma$ and $\pi_3$ numerically in the absence of nucleons -- but with ``tilt'' -- and found that for small $\epsilon$ the solution resembles the ``shifted CDW'' of Ref.\ \cite{Takeda:2018ldi}, but assumes more irregular shapes as $\epsilon$ is increased.}. This effect is ignored by working with the simple CDW ansatz, and we will minimize the {\it spatially averaged} free energy with respect to $\phi$ and $\vec{q}$ rather than attempting to work with the spatially nontrivial solution of the Euler-Lagrange equation. (Let alone attempting to find a self-consistent solution in the presence of the nucleons.) For convenience, we 
already introduce the spatial averaging on the level of the mean-field Lagrangian. After separating the $\vec{q}=0$ contribution, this amounts to replacing 
\begin{equation} \label{epsphi}
\epsilon\phi[1-\cos(2\vec{q}\cdot\vec{x})] \to \frac{\epsilon\phi}{V}\int d^3\vec{x} \,[1-\cos(2\vec{q}\cdot\vec{x})] = \epsilon\phi\left[1-\frac{\sin(qL_q)}{qL_q} \right] \to  \epsilon\phi \, ,
\end{equation}
where $V$ is the volume of the system and $L_q$ its length in the direction of $\vec{q}$. In the last step, we have taken $L_q\to \infty$ at fixed nonzero wave number $q$ to obtain the contribution to $\Delta U$ given in Eq.\ (\ref{UDelU}). We see that the result does not depend on $q$; in particular, taking the limit $q\to 0$ now does not change the contribution. If, on the other hand, we are interested in the isotropic case, we first let $q\to 0$ and then take the thermodynamic limit $L_q\to \infty$, in which case the  contribution (\ref{epsphi}) vanishes. This $q\to 0$ discontinuity results in the prefactor $1-\delta_{0q}$ in Eq.\ (\ref{UDelU}). It implies that it is energetically more costly by a finite amount to have an infinitesimally small winding per unit length (i.e., wavelength going to infinity) compared to the constant zero-winding solution. Again, this is a consequence of the explicit symmetry breaking and our use of the CDW ansatz; the discontinuity is absent in the chiral limit $\epsilon =0$.

\subsection{Free energy}
\label{sec:free}

The (yet to be renormalized) free energy from the Lagrangian (\ref{LMF}) is 
\be
\Omega = \Omega_{\rm bar} - \frac{m_\omega^2}{2}{\omega}^2-\frac{d}{4}{\omega}^4  + U + \Delta U \, ,
\ee
with the baryonic contribution $\Omega_{\rm bar}$, which is derived as follows. 

We first observe that the nucleonic sector of the mean-field Lagrangian (\ref{LMF}) is formally equivalent to a Lagrangian of free fermions. The thermodynamics can thus be straightforwardly computed without further approximations. To this end, we need to compute the fermionic spectrum in the presence of the CDW. We first identify the inverse nucleon propagator in momentum space,
\begin{equation}
S^{-1}(K)=-\gamma^\mu K_\mu + M -\mu_*\gamma^0+\vec{q}\cdot\vec{\gamma}\gamma^5\tau_3 \, , 
\end{equation}
where $K=(k_0,\vec{k})$, and $k_0=-i\omega_n$ with the fermionic Matsubara frequencies $\omega_n = (2n+1)\pi T$, where $T$ is the temperature and $n\in \mathbb{Z}$. The poles of the propagator $S(K)$ are given by the zeros of the determinant of $S^{-1}(K)$, which can be factorized as follows, 
\begin{equation}
\text{det}\, S^{-1} = [(k_0+\mu_*)^2-(E_k^+)^2]^2\,[(k_0+\mu_*)^2-(E_k^-)^2]^2 \, , 
\end{equation}
with the single-nucleon energies 
\begin{equation}\label{eq:dispersion}
E_k^\pm = \sqrt{\left(\sqrt{k_\ell^2+M^2}\pm q\right)^2+k_\perp^2} \, .
\end{equation} 
Here we have introduced longitudinal and transverse components of the single-particle momentum $\vec{k}$ with respect to the direction of the CDW, $\vec{k}_{\ell} = \hat{\vec{q}}\,\hat{\vec{q}}\cdot\vec{k}$, $\vec{k}_\perp = \vec{k}-\hat{\vec{q}}\,\hat{\vec{q}}\cdot\vec{k}$. We see from Eq.\ (\ref{eq:dispersion}) that the wave vector $\vec{q}$ introduces two different dispersion relations which would otherwise be degenerate. In our model, the dispersions have a very simple analytical form despite the presence of the CDW. This will be particularly useful for the regularization of the vacuum contribution, which can be done analytically. This is in contrast to the extended linear sigma model of Ref.\ \cite{Heinz:2013hza}, where the dispersions are complicated solutions of a quartic polynomial. 

Now, following the standard procedure of thermal field theory, we can compute the free energy density from the logarithm of the partition function. After performing the sum over Matsubara frequencies, we obtain the baryonic contribution 
\begin{equation}
\Omega_{\rm bar} = -2\sum_{e=\pm}\sum_{s=\pm} \int\frac{d^3\vec{k}}{(2\pi)^3}\left\{\frac{E_k^s}{2}+T\ln\left[1+e^{-(E_k^s-e\mu_*)/T}\right]\right\} \, , 
\end{equation}
where the prefactor 2 indicates the degeneracy of neutrons and protons. The free energy density $\Omega$ does not depend on space and thus all thermodynamic quantities will be homogeneous as well. This reflects the fact that translation symmetry is unbroken by the CDW -- at least in the chiral limit and within our approximation also in the presence of a nonzero pion mass. Of course, $\Omega$ {\it does} depend on the vector $\vec{q}$ and thus the anisotropy does show up in physical observables.

In all our results we restrict ourselves to zero temperature. In this case, with $\mu_*>0$, there is no anti-particle matter, i.e., the logarithm is only nonzero for $e=+1$. We obtain 
\begin{align} \label{Ombar}
\Omega_{\rm bar}  =-2(P_{\rm vac}+P_{\rm mat}) \, , 
\end{align}
with the (divergent) vacuum pressure of a single fermionic degree of freedom,
\begin{equation} \label{D0def}
P_{\rm vac}  \equiv \frac{1}{2\pi^2}\sum_{s=\pm} \int_0^\infty dk_\ell\int_0^\infty dk_\perp k_\perp E_k^s \, , 
\end{equation}
and the corresponding (finite) matter part, 
\bea \label{Pmatdef}
P_{\rm mat}  \equiv \frac{1}{2\pi^2}\sum_{s=\pm} \int_0^\infty dk_\ell\int_0^\infty dk_\perp k_\perp (\mu_*-E_k^s)\Theta(\mu_*-E_k^s) \, .
\eea
We have written the momentum integral in cylindrical coordinates and employed invariance of the integrand under $k_\ell\to -k_\ell$.

The double integral in the matter part can be evaluated analytically. After some tedious algebra due to the step function one can write the result as 
\bea \label{calD} \allowdisplaybreaks
P_{\rm mat}&=&\frac{\Theta(\mu_*-q-M)}{16\pi^2}\left\{M^2[M^2+4q(q-\mu_*)]\ln\frac{\mu_*-q+k_-}{M}+\frac{k_-}{3}[2(\mu_*^2-q^2)(\mu_*-q)-M^2(5\mu_*-13q)]\right\} \non[2ex]
&&+\frac{\Theta(\mu_*+q-M)}{16\pi^2}\left\{M^2[M^2+4q(q+\mu_*)]\ln\frac{\mu_*+q+k_+}{M}+\frac{k_+}{3}[2(\mu_*^2-q^2)(\mu_*+q)-M^2(5\mu_*+13q)]\right\} \non[2ex]
&&+\frac{\Theta(q-\mu_*-M)}{16\pi^2}\left\{M^2[M^2+4q(q-\mu_*)]\ln\frac{q-\mu_*+k_-}{M}-\frac{k_-}{3}[2(\mu_*^2-q^2)(\mu_*-q)-M^2(5\mu_*-13q)]\right\} \non[2ex]
&&-\frac{\Theta(q-M)}{8\pi^2}\left[M^2(M^2+4q^2)\ln\frac{q+\sqrt{q^2-M^2}}{M}-\frac{\sqrt{q^2-M^2}}{3}q(2q^2+13M^2)\right] \, ,
\eea
where we have abbreviated 
\begin{equation} \label{kpm}
k_\pm\equiv \sqrt{(\mu_*\pm q)^2-M^2} \, .
\end{equation}
As a check, one finds for $q=0$, 
\be
P_{\rm mat} = \frac{\Theta(\mu_*-M)}{8\pi^2}\left[\mu_* k_F \left(\frac{2}{3}k_F^2-M^2\right)+M^4\ln\frac{\mu_*+k_F}{M}\right]
\, , 
\ee
where $k_F=\sqrt{\mu_*^2 - M^2}$ is the Fermi momentum. This is the zero-temperature pressure of a non-interacting fermion gas with chemical potential $\mu_*$ and fermion mass $M$. Moreover, for $M=0$ we find 
\be\label{PM0}
P_{\rm mat} =\frac{\mu_*^4}{12\pi^2} \, , 
\ee
which is the pressure of massless fermions. In particular, the wave number $q$ has dropped out. This is expected since any modulation is irrelevant if the amplitude, here $M$, is zero.

The Dirac sea contribution $P_{\rm vac}$ has to be treated more carefully. We explain all details in Appendix \ref{sec:diracsea} and proceed here with a short summary and the final result for the renormalized free energy. We first employ proper time regularization to identify the divergent contributions, which we can express in terms of a proper time cutoff $\Lambda$ and a renormalization scale $\ell$. Then, we renormalize our model by introducing the renormalized field $\phi_r$ and renormalized parameters $f_{\pi,r}, g_{\sigma,r}, a_{n,r}, \epsilon_r$. They are related to the bare quantities of the original Lagrangian by counterterms $\delta a_n$ and a field rescaling factor $Z$ for the sigma and pion fields. The divergent parts of $\delta a_n$ and $Z$ are fixed to cancel the divergences of $P_{\rm vac}$. The $q=0$ part of the Dirac sea is uniquely determined by our fit of the model parameters to properties of nuclear matter, and no dependence on the choice of the finite parts of $\delta a_n$, $Z$, and on the scale $\ell$ is left. The $q$-dependent part, however, is less straightforward, and we keep the scale $\ell$ in the following results to discuss our choice for it carefully. Dropping for notational convenience the subscript $r$ from the renormalized quantities, the calculation in Appendix \ref{sec:diracsea} yields
\be \label{Omren}
\Omega =  - 2P_{\rm mat} - \frac{m_\omega^2}{2}{\omega}^2-\frac{d}{4}{\omega}^4  + \tilde{U} + \Delta \tilde{U}  \, . 
\ee
The renormalized Dirac sea contribution is absorbed in the modified 
contributions to the meson potential, 
\begin{subequations}
\bea \label{Utilde}
\tilde{U} &\equiv& \tilde{U}(\phi) \equiv U(\phi)+\frac{m_N^4}{96\pi^2}(1-8\varphi^2-12\varphi^4\ln\varphi^2 +8\varphi^6 -\varphi^8) \, , \\[2ex]
\label{DelUtilde}
\Delta\tilde{U} &\equiv& \Delta\tilde{U}(\phi,q) \equiv \Delta U(\phi,q)-\frac{q^2M^2}{2\pi^2}\ln\frac{M^2}{\ell^2} -\frac{q^4}{2\pi^2}F(y)\, ,
\eea
\end{subequations}
where
\be \label{varphidef}
\varphi\equiv\frac{\phi}{f_\pi} =  \frac{M}{m_N} \, , 
\ee
with the nucleon mass in the vacuum $m_N=939\, {\rm MeV}$, and 
\begin{equation}\label{Fdef}
F(y) \equiv \frac{1}{3}+\Theta(1-y)\left[-\sqrt{1-y^2}\frac{2+13y^2}{6}+2y^2\left(1+\frac{y^2}{4}\right)\ln\frac{1+\sqrt{1-y^2}}{y} \right] \, ,
 \end{equation}
 with 
 \be \label{ydef}
y\equiv \frac{M}{q} \, . 
\ee
The terms in $\tilde{U}$ generated by the nucleonic Dirac sea are of order $(\phi^2-f_\pi^2)^5\sim(\varphi^2-1)^5$ and higher. The reason is that we fit all coefficients in front of $(\varphi^2-1)^n$ with $n\le 4$ to reproduce physical quantities, and thus the corrections by the Dirac sea to the terms up to order $ (\varphi^2-1)^4$ are absorbed by the fit.  

 The $q$-dependent contribution of the Dirac sea in $\Delta\tilde{U}$ contains the renormalization scale $\ell$. Let us discuss two limits that will serve us to choose $\ell$. Firstly, in the vacuum, where  $\omega=P_{\rm mat}=0$, $\phi=f_\pi$ (i.e., $M=m_N$), we find in the chiral limit, where $\epsilon=0$, 
\be
\mbox{vacuum:} \qquad \Omega = 2f_\pi^2q^2\left(1-\frac{g_\sigma^2}{4\pi^2}\ln\frac{m_N^2}{\ell^2}\right)-\frac{q^4}{6\pi^2} \, ,
\ee
where we have assumed $q<m_N$, such that the step function in Eq.\ (\ref{Fdef}) vanishes. Secondly, in the limit of large $q$, with all other quantities kept finite, we have 
\be
\mbox{large $q$:} \qquad \Omega = \frac{q^2M^2}{2\pi^2}\left(2+\frac{4\pi^2}{g_\sigma^2}-\ln\frac{4q^2}{\ell^2}\right) + {\cal O}(q^0) \, .
\ee
[$F(y)$ contributes to the logarithm and the matter part reduces to the limit (\ref{PM0}) and thus is subleading.] We now require that for small $q$ in the vacuum  $\Omega = 2f_\pi^2q^2 +{\cal O}(q^4)$ \cite{Broniowski:1990gb,Carignano:2014jla} and that the free energy be bounded from below as $q\to \infty$. Consequently, a natural, albeit not unique, choice is  
\be \label{ellchoice}
\ell = \sqrt{m_N^2 +(2q)^2} \, .
\ee
The $q$-dependence is crucial to avoid the unboundedness of the free energy, which was identified as a problem in previous works in similar models \cite{Broniowski:1990gb,Carignano:2014jla}. Since $q$ will be determined dynamically as we vary the chemical potential, the scale $\ell$ depends on the medium. This is typical for perturbative calculations in renormalizable theories such as QCD, if applied to nonzero temperatures and/or densities \cite{kapustabook,Kurkela:2009gj,Fraga:2023cef}. At the end of Sec.\ \ref{sec:phase_structure}, we shall further discuss the choice (\ref{ellchoice}) by comparing our results to the ones obtained with  $\ell=m_N$, where $\Omega$ {\it is} unbounded from below. 

\subsection{Stationarity equations}
\label{sec:stat}

The thermodynamically stable phases are determined  by minimizing the renormalized free energy with respect to the condensates $\phi$, $\omega$, and the wave number $q$,
\be \label{dOm}
\frac{\partial\Omega}{\partial \phi} = \frac{\partial\Omega}{\partial \omega} = \frac{\partial\Omega}{\partial q} = 0 \, .
\ee
All derivatives are taken with the two other dynamical quantities, the chemical potential, and the scale $\ell$ kept fixed. The minimization with respect to $q$ is equivalent to requiring the total axial current to vanish; for a nonzero $q$ this means that there will be counter-propagating currents from the mesonic and the baryonic sectors which cancel each other. More explicitly, the stationarity equations (\ref{dOm}) read
\begin{subequations}\label{stat}
\bea
g_\sigma n_s &=& -\tilde{U}'(\phi)-4q^2\phi\left[1-\frac{g_\sigma^2}{4\pi^2}\left(1+\ln\frac{M^2}{\ell^2}\right)\right]-(1-\delta_{0q})\epsilon+ \frac{g_\sigma q^3}{2\pi^2}F'(y) 
 \, , \label{eq:sigma} \\[2ex]
g_\omega n_B&=&m_\omega^2\omega+d\omega^3  \, , \label{eq:omega} \\[2ex]
j &=& -4q\phi^2\left(1-\frac{g_\sigma^2}{4\pi^2}\ln\frac{M^2}{\ell^2}\right)+\frac{q^3}{2\pi^2}[4F(y)-yF'(y)] \, , \label{eq:q}
\eea
\end{subequations}
where we have introduced the scalar density $n_s$, the baryon density $n_B$, and the contribution to the axial current from the baryons $j$,
\be
n_s = -2\frac{\partial P_{\rm mat}}{\partial M} \, , \qquad n_B = 2\frac{\partial P_{\rm mat}}{\partial \mu} \, , \qquad j = -2\frac{\partial P_{\rm mat}}{\partial q} \, .
\ee
These quantities are computed straightforwardly from the expression (\ref{calD}). For completeness, and for a brief discussion of their limits, we present their explicit expressions in Appendix \ref{ap:functions}.

\subsection{Fitting parameters}
\label{sec:fit}

Since we fit our parameters to vacuum properties and the properties of nuclear matter in the absence of the CDW, the matching procedure is very similar to our previous works within the same model  \cite{Fraga:2018cvr,Schmitt:2020tac,Fraga:2022yls}. Due to  empirical uncertainties and in order to explore all qualitatively different scenarios of the model, we shall not simply work with a single parameter set but explore the parameter space of the model within and somewhat beyond these uncertainties. It is therefore 
useful to explain the details of our parameter fitting. 

We first fit $g_\sigma$ from the vacuum mass of the nucleon, $m_N = g_\sigma f_\pi$, where we have used that in the vacuum $\phi=f_\pi$. Next, we compute the pion and sigma masses by temporarily reinstating mesonic fluctuations about the $q=0$ vacuum. This can be done for instance by replacing $\phi^2\to (f_\pi+\sigma)^2+\pi^2$ in the potential $\tilde{U}$ from Eq.\ (\ref{Utilde}) and expanding in the fluctuations $\sigma$ and $\pi$. The coefficients in front of the quadratic terms $\pi^2/2$, $\sigma^2/2$ yield the masses squared, 
\bea \label{mpimsig}
m_\pi^2  = \frac{\tilde{U}'(f_\pi)+\epsilon}{f_\pi} = a_1 \, , \qquad 
m_\sigma^2 = \tilde{U}''(f_\pi) = m_\pi^2+f_\pi^2 a_2 \, . 
\eea
The first relation is used to fix $a_1$ from the pion mass $m_\pi$.
In our results, we shall consider both the chiral limit $m_\pi=0$ and the physical case $m_\pi=139\, {\rm MeV}$. Requiring that $\phi=f_\pi$ satisfy \eqref{eq:sigma} in the vacuum then gives $\epsilon = f_\pi m_\pi^2$. Due to the very uncertain (and not uniquely defined) physical value of $m_\sigma$ we shall not use the second relation to fix $a_2$, but use this relation to compute $m_\sigma$ once $a_2$ is fixed from other constraints, which is useful for a comparison to other models.

The remaining parameters are $g_\omega$, $a_2$, $a_3$, $a_4$, and $d$. We relate them to the following properties of isospin-symmetric nuclear matter at saturation: 
the binding energy $E_B=-16.3$ MeV, leading to a chemical potential for the baryon onset $\mu_0\equiv m_N + E_B= 922.7$ MeV, the baryon density  $n_0=0.153 \, {\rm fm}^{-3}$, the effective Dirac mass  $M_0 \simeq (0.7 - 0.8) m_N$, and the incompressibility $K \simeq (200-300)\, {\rm MeV}$. Following Ref.\ \cite{Fraga:2022yls}, we denote the solution of Eq.\ \eqref{eq:omega} at $n_B=n_0$ by 
\begin{equation}
\omega_0= \frac{g_{\omega}n_0}{m_\omega^2}f(x_0)\, , 
\end{equation}
with 
\begin{equation}
f(x) \equiv \frac{3}{2x}\frac{1-(\sqrt{1+x^2}-x)^{2/3}}{(\sqrt{1+x^2}-x)^{1/3}} \, , \qquad x_0\equiv \frac{3\sqrt{3d}\,g_{\omega}n_0}{2m_\omega^3} \, . 
\end{equation}
The effective chemical potential at saturation is $\mu_0^* = \mu_0-g_\omega \omega_0$. Inserting this into Eq.\ \eqref{eq:omega} gives a quadratic equation for $g_\omega^2$ with the relevant solution 
\begin{equation} \label{gomd}
g_{\omega}^2 = \frac{m_\omega^2}{2n_0}(\mu_0-\mu_0^*)\left[1+\sqrt{1+\frac{4dn_0(\mu_0-\mu_0^*)}{m_\omega^4}}\right] \, .
\end{equation}
Since the effective chemical potential can also be written as $\mu_0^*= \sqrt{k_F^2+M_0^2}$ with the Fermi momentum at saturation $k_F = (3\pi^2n_0/2)^{1/3}$, Eq.\ (\ref{gomd})  is a relation between the model parameters $g_\omega$ and $d$, all other quantities being physical parameters whose values can be inserted later. We see that at $\mu_0=\mu_0^*$ the coupling $g_\omega$ vanishes, which translates into an upper bound for $M_0$, 
\be \label{M0limit}
M_0 < \sqrt{\mu_0^2-k_F^2} \simeq 0.943\, m_N \, .
\ee
For the remaining parameters $a_2, a_3, a_4$ we set up the following three coupled equations: the definition of the incompressibility $K$ (see for instance Ref.\ \cite{Schmitt:2020tac} for bringing the definition into the form used here), the free energy of saturated nuclear matter being equal to that of the vacuum (here 0), and the stationarity equation for $\phi$ \eqref{eq:sigma},
\begin{subequations} \label{a1a2a3fit}
\bea
0 &= & \tilde{U}''(\phi)+\frac{g_{\sigma}^2}{\pi^2}\left(\frac{k_F^3+3k_FM_0^2}{\mu_0^*}-3M_0^2\ln\frac{\mu_0^*+k_F}{M_0}\right)
+\frac{\displaystyle{\frac{6g_{\sigma}^2k_F^3}{\pi^2}\left(\frac{M_0}{\mu_0^*}\right)^2}}{\displaystyle{K-\frac{6k_F^3}{\pi^2}\frac{g_{\omega}^2}{m_\omega^2}[f(x_0)+x_0f'(x_0)]
-\frac{3k_F^2}{\mu_0^*}}} \, ,
\\[2ex]
0&= & \frac{m_\omega^2}{2}\omega_0^2+\frac{d}{4}\omega_0^4-\tilde{U}(\phi)+\frac{1}{4\pi^2}\left[\left(\frac{2}{3}k_F^3-M_0^2k_F\right)\mu_0^*+M_0^4\ln\frac{k_F+\mu_0^*}{M_0}\right] \, , \label{Psat}\\[2ex]
0&=&  \tilde{U}'(\phi)+ \frac{g_{\sigma}M_0}{\pi^2}\left(k_F\mu_0^*-M_0^2\ln\frac{k_F+\mu_0^*}{M_0}\right) \, ,\label{sigsat}
\eea
\end{subequations}
where the potential $\tilde{U}$ and its derivatives are evaluated at saturation, $\phi=M_0/g_\sigma$. The parameters $a_2, a_3, a_4$ only appear in $\tilde{U}$ and its derivatives. Hence, despite their tedious look, Eqs.\ (\ref{a1a2a3fit}) form a simple system of linear equations for these parameters. As a consequence, we can derive elementary (but very lengthy) analytical expressions for $a_2, a_3, a_4$ purely in terms of physical quantities and the model parameter $d$.  

In our results we shall consider different values of the quartic vector meson coupling $d$ while keeping the aforementioned properties of symmetric nuclear matter at saturation fixed. In order to translate this coupling constant into a more physical quantity, we temporarily consider isospin-asymmetric matter with a Yukawa coupling $g_{\rho}$ between the nucleons and the rho meson. This allows us to relate our parameters to the symmetry energy $S\simeq (30-34)\, {\rm MeV}$ and the ``slope parameter'' $L$, which characterizes the change of the symmetry energy under variation of the baryon number. For the exact definitions of $S$ and $L$ see for instance Ref.\ \cite{Fraga:2022yls}, from which we also quote the relevant relations 
\begin{subequations}
\bea
g_{\rho}^2 &=& \frac{3\pi^2m_\omega^2}{k_F^3}\left(S-\frac{k_F^2}{6\mu_0^*}\right)\left(1+\frac{d\omega_0^2}{m_\omega^2}\right) \, ,\label{eq:gNrho}\\[2ex]
L &=& \frac{3g_{\rho}^2n_0}{2(m_\rho^2+d\omega_0^2)}\left[1-\frac{2d\,n_0g_{\omega}\omega_0}{(m_\rho^2+d\omega_0^2)(m_\omega^2+3d\omega_0^2)}\right]+\frac{k_F^2}{3\mu_0^*}\left(1-\frac{K}{6\mu_0^*}\right)+\frac{g_{\omega}^2n_0k_F^2}{2m_\omega^2\mu_0^{*2}}[f(x_0)+x_0f'(x_0)] \, ,\label{eq:L}
\eea
\end{subequations}
with the rho meson mass $m_\rho\simeq 776\, {\rm MeV}$. 
We shall work with $S=32\, {\rm MeV}$, such that these equations give us a map between $d$ and $L$ if all other model parameters are fixed. The value of $L$ is poorly known, with experimental data indicating a range $L\simeq (40-140)\, {\rm MeV}$ \cite{Lattimer:2012xj,Oertel:2016bki,Tews:2016jhi,PREX:2021umo,Reed:2021nqk,Sotani:2022hhq}, which is not violated for any $d$ considered in this paper. 

We summarize our fitting procedure as follows: $m_\omega$ and $g_\sigma$ are fixed in all cases we consider, and the value of $m_\pi$ fixes $a_1$ and $\epsilon$; then, the parameters $g_\omega,a_2,a_3,a_4,d$ are determined from $\mu_0,n_0,M_0,K,L$, where $\mu_0,n_0$ are always taken to assume their well-known values, while we will explore the dependence on the less well known $M_0, K, L$.

\section{Results}\label{sec:results}

\subsection{Isotropic matter: absence of first-order transition due to Dirac sea}
\label{sec:isotropic}

\begin{figure}
\includegraphics[width=0.495\linewidth]{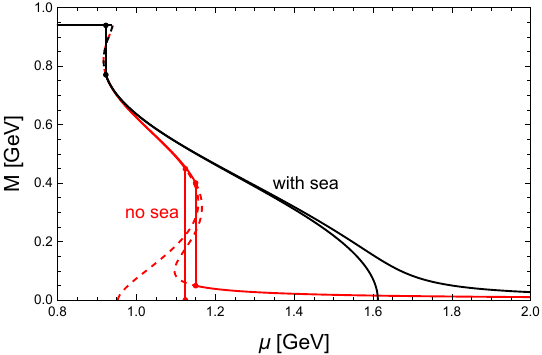}
\includegraphics[width=0.495\linewidth]{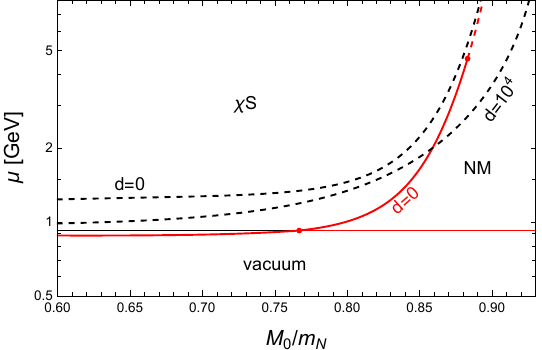}
\caption{Effect of the nucleonic vacuum contribution (``Dirac sea'') on the chiral phase transition. \textit{Left panel:} Effective nucleon mass $M$ as a function of the baryon chemical potential $\mu$ with (black) and without (red) Dirac sea. Solid lines represent stable phases, while dashed lines indicate metastable or unstable solutions. For each color we display the chiral limit (curve connects to the $M = 0$ solution) and the physical case  (curve asymptotes to $M = 0$). First-order transitions are marked by solid vertical lines. The baryon onset (small vertical lines at $\mu=\mu_0$) is the same in all cases by construction. The parameters for this panel are $M_0=0.82\, m_N$, $K=250\, {\rm MeV}$, $d=0$. 
\textit{Right panel:} Phase transitions in the chiral limit upon variation of $M_0$, keeping $K=250\, {\rm MeV}$, for $d=0$ (with and without Dirac sea) and $d=10^4$ (for the case with Dirac sea). Colors as in the left panel; solid (dashed) lines are first- (second-) order transitions between vacuum, nuclear matter (NM) and the chirally restored phase ($\chi$S). In the case with Dirac sea, the baryon onset occurs at $\mu=\mu_0$ for all $M_0$ and $d$ (solid black line, partly covered by the red line).  }
\label{fig:isotropic}
\end{figure}

To lay the ground for the discussion of the CDW we first focus on the isotropic case $q=0$. 
For given $M_0$, $K$, and $d$ we can solve the stationarity equations (\ref{eq:sigma}), (\ref{eq:omega}) for $\phi$, $\omega$ as  functions of $\mu$ and insert the result back into Eq.\ (\ref{Omren}) to compute the corresponding free energy. [The stationarity equation (\ref{eq:q}) is trivially solved by $q=0$.] Here, in the isotropic case, the results do not depend on the renormalization scale $\ell$. The result for  the effective baryon mass $M$ is shown in the left panel of Fig.\ \ref{fig:isotropic}, where we have included four cases: with/without Dirac sea and zero/physical pion mass. The specific values for the model parameters needed to compute these results are given in Table \ref{table:para} in Appendix \ref{ap:para}. We see that in the no-sea approximation the chiral transition is of first order, for either value of the pion mass. The critical chemical potential of the first-order chiral transition is determined by finding the point where the free energies 
of chirally broken and chirally restored matter are equal.  Including the Dirac sea renders the transition second order (chiral limit) or turns it into a crossover (physical pion mass), and moves it to significantly larger values of $\mu$, in accordance with Ref.\ \cite{Brandes:2021pti}. 

Does this observation depend on the specific parameter choice? This question is addressed in the right panel of Fig.\ \ref{fig:isotropic}, where we explore the behavior of the chiral phase transition as a function of the parameter $M_0$. In this plot we restrict ourselves to the chiral limit to avoid any crossovers, which are difficult to display due to the absence of a well-defined critical chemical potential. For any $M_0$ we adjust the model parameters such that $K=250\, {\rm MeV}$ is held fixed (as well as all other properties of symmetric nuclear matter discussed in Sec.\ \ref{sec:fit}). We see that in the no-sea approximation there is a region of small $M_0$ --  in fact covering a large part of the physically most likely values of $M_0$ -- where there is a direct transition from the vacuum to chirally symmetric matter. This means chirally symmetric matter is stable at zero pressure, which is reminiscent of the strange quark matter hypothesis \cite{PhysRevD.4.1601,PhysRevD.30.272}. This interpretation is somewhat far fetched in the current approach but becomes more sensible if strangeness is included, where indeed this behavior persists \cite{Fraga:2022yls}. In an intermediate range of $M_0$ we observe a baryon onset from the vacuum to nuclear matter, followed by a first-order chiral transition. This is the scenario of the left panel. Then, for even larger values of $M_0$ the chiral transition becomes second order (and moves to extremely large $\mu$) even in the no-sea approximation. 
In the presence of the Dirac sea, the behavior is qualitatively the same for all values of $M_0$: The first-order baryon onset at $\mu=\mu_0$ is followed by a second-order chiral transition, shown by the black dashed curves. Even a large mesonic self-coupling $d=10^4$ does not change this conclusion. We have also varied the incompressibility $K$ within the empirically preferred regime (not shown in the plot) and never found a first-order transition when the Dirac sea is included.

What does this imply for the case of a physical pion mass? The left panel of Fig.\ \ref{fig:isotropic} shows how the second-order transition becomes a crossover if the pion mass is switched on.  Therefore, the result of the right panel indicates that in the presence of a physical pion mass the chiral transition is always a crossover (assuming isotropy). This is the basis on which we now investigate the CDW.

\subsection{CDW solution}
\label{sec:res_chiral_limit}

We will now stick to the full calculation that takes into account the Dirac sea and only comment on the differences to the no-sea approximation in Sec.\ \ref{sec:compare}. To find $q> 0$ solutions to the stationarity equations \eqref{stat} it is useful to start with the chiral limit. In this case, the CDW branch can connect continuously to the isotropic solution. The first possibility to connect is with the $q=0$ nuclear matter branch. The chemical potential where the CDW attaches to this branch can be computed from the $q\to 0$ limit of Eq.\ (\ref{eq:q}), 
\bea \label{attach1}
\ell \exp\left(\frac{2\pi^2}{g_\sigma^2}\right) = M+\Theta(\mu_*-M)(\mu_*-M+\sqrt{\mu_*^2-M^2}) \, ,
\eea
where $M$ and $\omega$ (hidden in $\mu_*$) are computed from Eqs.\ (\ref{eq:sigma}) and (\ref{eq:omega}) with $q=0$. This equation describes the appearance of the CDW via the infinite-wavelength limit at finite amplitude of the chiral condensate. 
Secondly, the CDW can connect continuously to the chiral solution $M=0$. The corresponding $q$ and $\mu$ can be computed from the 
$M\to 0$ limit of Eqs.\ (\ref{eq:sigma}) and (\ref{eq:q}), see also Appendix \ref{ap:functions},
\begin{subequations} \label{ch12}
\bea
\frac{\pi^2}{g_\sigma^2}\tilde{U}''(0) +\mu_*^2-\mu_*q\ln\left|\frac{\mu_*+q}{\mu_*-q}\right| &=& q^2\left(\ln\frac{4|\mu_*^2-q^2|}{\ell^2}-2-\frac{4\pi^2}{g_\sigma^2}\right) \, , \label{ch1}\\[2ex]
-\frac{\mu_*q}{2}\ln\left|\frac{\mu_*+q}{\mu_*-q}\right| &=& q^2\left(\ln\frac{4|\mu_*^2-q^2|}{\ell^2} -1-\frac{4\pi^2}{g_\sigma^2}\right) \label{ch2} \, , 
\eea
\end{subequations}
where $\omega$ (hidden in $\mu_*$) is computed from Eq.\ (\ref{eq:omega}) at $M=0$. These equations describe the appearance of the chiral condensate from the zero-amplitude limit with a finite CDW wavelength.  Both Eqs.\ (\ref{attach1}) and (\ref{ch12}) connect the CDW to phases of nonzero baryon density. There is a third option for the CDW branch to end, namely in a $q\neq 0$ vacuum. These exotic vacua, in which the chiral condensate is anisotropic and the nucleons only contribute through the Dirac sea, but not via a nonzero density, play no role for the actual phase structure, as they are thermodynamically disfavored.

\begin{figure}
\begin{center}
\hbox{
\includegraphics[width=0.33\textwidth]{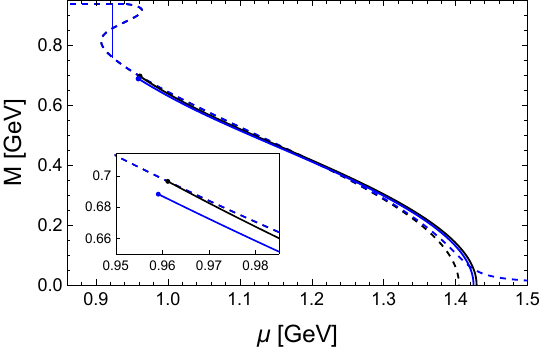}
\includegraphics[width=0.33\textwidth]{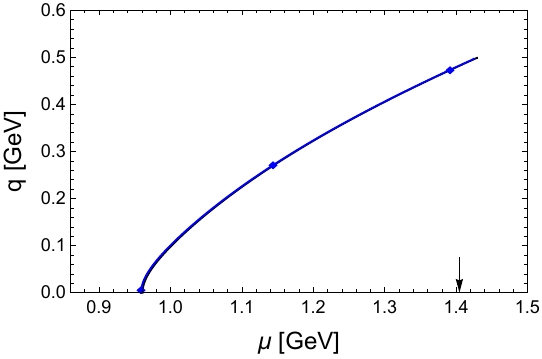}
\includegraphics[width=0.33\textwidth]{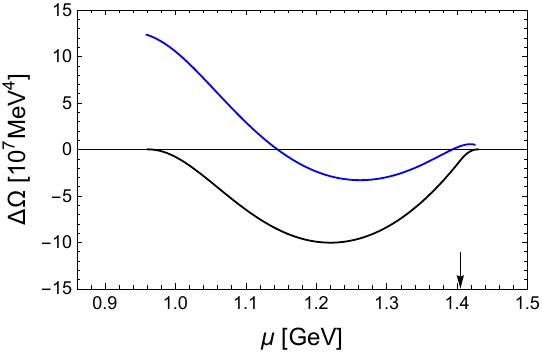}
}
\caption{Effective nucleon mass $M$ (left), wave number $q$ (middle), and free energy of the CDW phase minus that of the thermodynamically stable isotropic phase $\Delta\Omega$ (right),  for the chiral limit (black) and the physical pion mass (blue), as functions of the baryon chemical potential $\mu$. The parameters used are $M_0=0.81 m_N$, $K=250\, {\rm MeV}$, $d=10^4$. The arrow indicates the chiral phase transition in the chiral limit (in the chiral limit, $\Delta\Omega$ is measured relative to the chirally broken phase to the left of the arrow and relative to the chirally restored phase to the right of the arrow). The dashed lines in the left panel are the $q=0$ curves, including the baryon onset shown as a vertical solid line, while the three markers in the middle panel indicate the points for which we show the nucleonic Fermi surfaces in Fig.\ \ref{fig:fermi_surfaces}. }
\label{fig:specific_case}
\end{center}
\end{figure}

\begin{figure}
\begin{center}
\includegraphics[height=0.23\textwidth]{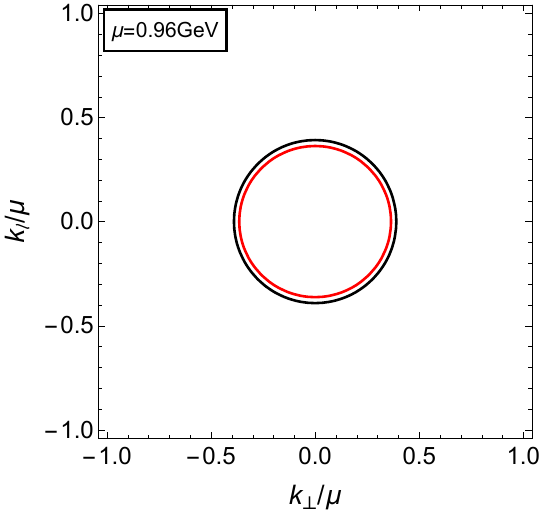}\hspace{0.5cm}
\includegraphics[height=0.23\textwidth]{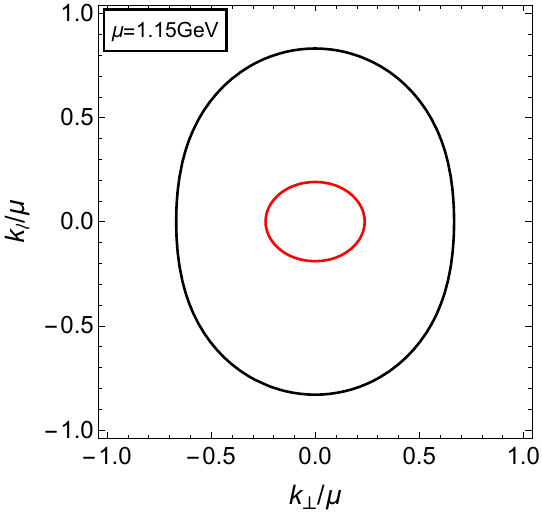}\hspace{0.5cm}\includegraphics[height=0.23\textwidth]{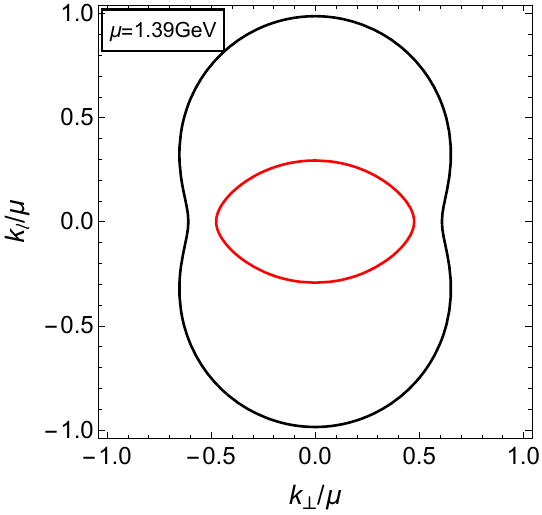}
\end{center}
\caption{Fermi surfaces $\mu_*=E_k^+$ (red) and $\mu_*=E_k^-$ (black) in the plane of longitudinal and transverse momentum components with respect to $\vec{q}$, such that the actual 2-dimensional Fermi surfaces are obtained by rotation around the $k_\perp=0$ axis. The three plots correspond to three different chemical potentials, as indicated in the middle panel of Fig.~\ref{fig:specific_case}, and are calculated with the physical pion mass (and $M_0$, $K$, $d$ as in Fig.\ \ref{fig:specific_case}). }
\label{fig:fermi_surfaces}
\end{figure}

We start by discussing the CDW for a specific parameter set with fixed values of $M_0$, $K$, $d$, and the scale $\ell$ given by Eq.\ (\ref{ellchoice}). Again, for the precise model parameters used here, see Table \ref{table:para} in Appendix \ref{ap:para}. The numerical solutions are shown in Figs.\ \ref{fig:specific_case} and \ref{fig:fermi_surfaces}. The main observations are as follows.

\begin{itemize}
\item {\it Chiral limit.} The black curves in Fig.\ \ref{fig:specific_case} show that the second-order isotropic chiral phase transition is replaced by two transitions at $\mu\simeq 0.96\, {\rm GeV}$  and $\mu\simeq 1.43\, {\rm GeV}$ between which the CDW is energetically favored. The lower end of this region is a nonzero-amplitude, infinite-wavelength transition as described by Eq.\ (\ref{attach1}), while the upper end of this region is a zero-amplitude, finite-wavelength transition as described by Eqs.\ (\ref{ch12}). The values of the effective mass $M$ (left panel of Fig.\ \ref{fig:specific_case}) show that a CDW develops in nuclear matter, and as we move towards larger $\mu$ the effective mass decreases such that we observe a CDW in almost chirally symmetric matter.

\item {\it Effect of explicit chiral symmetry breaking.} The left panel (see zoom-in) and the right panel of Fig.\ \ref{fig:specific_case} show that in the case of a physical pion mass the CDW solution does not connect continuously to the isotropic branch. This is due to the term $\epsilon \phi$ in $\Delta U$ (\ref{UDelU}), whose origin is explained below that equation. In particular, the CDW solution admits $M\to 0$ even if $m_\pi$ is nonzero, although it becomes energetically disfavored at a nonzero $M$. As a consequence, the CDW region is now bounded by two first-order transitions and has shrunk, but not disappeared completely. The CDW exists although, in the absence of anisotropic phases, the chiral transition is a crossover, i.e., the crossover is disrupted by two phase transitions that break (entrance to the CDW) and restore (exit from the CDW) rotational symmetry.

\item {\it Fermi surfaces.} In Fig.\ \ref{fig:fermi_surfaces} we show the Fermi surfaces of the two fermion states $s=\pm$ given by the dispersion relations (\ref{eq:dispersion}). For each dispersion, all states in momentum space $(k_\ell,\vec{k}_\perp)$ are filled up to the Fermi surface defined by $\mu_*=E_k^s$, as indicated by the step function in Eq.\ (\ref{Pmatdef}). For given $q$, $M$ these Fermi surfaces can easily be computed, and Fig.\ \ref{fig:fermi_surfaces} shows them for 3 different chemical potentials, using the physical pion mass. The chemical potentials in the middle and right panels correspond to the two ends of the CDW region. More exotic topologies are possible -- disappearance of one of the Fermi surfaces (red) and split of the (black) Fermi surface into two disconnected regions -- but not realized here. Even though the Fermi surfaces are symmetric under $k_\ell\to -k_\ell$, there is a nonzero axial current in the vertical direction, to counterbalance that of the mesonic sector. The reason is that the two $s=\pm$ states contribute with opposite sign to that current, at least for $q<M$, as the integral in the first line of Eq.\ (\ref{dPdq}) shows. Therefore, for $q=0$, where red and black lines would be exactly on top of each other, no net fermionic current exists, while a net current starts to form for $q>0$ when the two Fermi surfaces no longer coincide. The case $q>M$ (realized in the right panel) is more complicated, because in this case the $s=-$ state has different regions in momentum space which contribute to the axial current with different sign, which again can be seen from the integrand in Eq.\ (\ref{dPdq}).   

\end{itemize}

\subsection{Locating the CDW in the parameter space}\label{sec:phase_structure}

Having discussed the details of a specific parameter choice, we now turn to a more systematic exploration of the parameter space of the model. This is necessary due to the large empirical uncertainties in particular of the quantities $M_0$, $K$, and $L$. Moreover, we have to keep in mind that our model is of phenomenological nature and we can, at best, make qualitative predictions and suggestions for QCD. Therefore, even regions at the edges or beyond the empirically allowed regions, which appear unlikely to be realized from the point of view of our model, may contain interesting features that are possibly relevant for QCD.

\begin{figure}
\begin{center}
\hbox{
\includegraphics[width=0.5\textwidth]{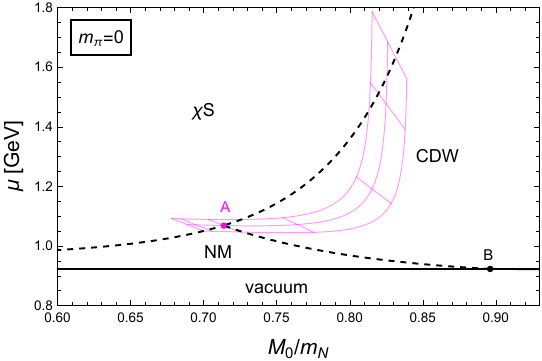}
\includegraphics[width=0.5\textwidth]{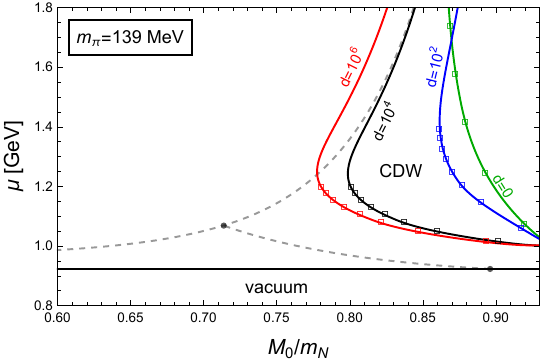}
}
\caption{{\it Left panel:} Zero-temperature phases in the chiral limit as the model parameter $M_0$ is varied, with fixed $K=250\, {\rm MeV}$, $d=10^4$  (NM -- isotropic nuclear matter, $\chi$S -- isotropic chirally restored phase, CDW -- chiral density wave). Black solid (dashed) lines are first (second) order phase transitions. The magenta grid indicates the variation of  the triple point A for $K=200,250,300\, {\rm MeV}$ (right to left) and $d=0,10^{1,2,3,4,5,6}$ (top to bottom). 
The triple point B
is always at the same $\mu$ and varies in $M_0$ by less than 0.5\%
as $K$ and $d$ are varied within the range given by the magenta grid. {\it Right panel:} Same as left panel, but now for the physical pion mass. The pale black lines are copied from the left panel for reference. Three  additional curves (red, blue, green) are shown for three different values of  $d$. The markers on the curves indicate  the baryon densities $n_B = (6,8,\ldots,20)n_0$ (from low to high $\mu$) on the isotropic side of the first-order phase transition. }
\label{fig:phase_diagram}
\end{center}
\end{figure}

The zero-temperature phase structure in the plane spanned by $\mu$ and the model parameter $M_0$ is shown in Fig.\ \ref{fig:phase_diagram}. Let us first discuss the chiral limit (left panel) and focus on the parameters $K=250\, {\rm MeV}$, $d=50$ (black curves). There are 3 qualitatively different scenarios. 

\begin{enumerate}

\item[$(i)$] For sufficiently small effective masses at saturation, $M_0\lesssim 0.71 m_N$, the chiral transition is unaffected by the CDW, there is a second-order transition between isotropic nuclear matter and the isotropic chirally restored phase. 

\item[$(ii)$] For $0.71\lesssim M_0/m_N \lesssim 0.90$ we find the scenario from Fig.\ \ref{fig:specific_case}: There is a finite CDW region covering the would-be isotropic chiral transition. 

\item[$(iii)$] As we approach the point B in the figure, the transition from isotropic nuclear matter to the CDW approaches the nuclear matter onset at $\mu_0$. For $M_0$ beyond that point, $M_0\gtrsim 0.90 m_N$, the model predicts a direct transition from the vacuum to the CDW. This transition occurs at an $M_0$-dependent critical chemical potential $\mu<\mu_0$, although on the scale of the plot the corresponding curve is indistinguishable from a horizontal line. Since we know that symmetric nuclear matter at saturation is isotropic, this appears to be an unphysical regime of our model. However, we need to keep in mind that the chiral limit of the left panel is unphysical anyway; and indeed, the right panel shows that this unphysical behavior is gone for the case of a physical pion mass. 

\end{enumerate}

How does the CDW region change as we vary the incompressibility $K$ and the meson coupling $d$? As a measure for the importance of the CDW we consider the range in $M_0$ between the points A and B, read off of the magenta grid (the point $B$ is essentially constant under the variations considered here). We vary $K$ within its empirically most likely range $K\simeq (200 - 300)\, {\rm MeV}$ and the quartic meson coupling $d=0-10^6$. For a connection to real-world quantities it is useful to translate the value of $d$ into the resulting slope parameter of the symmetry energy $L$ and also consider the corresponding sigma mass $m_\sigma$. This translation is shown in Fig.\ \ref{fig:para} in Appendix \ref{ap:para}. We read off for instance that for $K=250\, {\rm MeV}$ and tracking the location of point A as $d=0-10^6$, we obtain a range of $L\simeq (87 - 52)\, {\rm MeV}$ and $m_\sigma\simeq (720 - 830)\, {\rm MeV}$. The magenta grid thus illustrates for instance that the CDW becomes more important for increasing $K$ or increasing $d$ (decreasing $L$). Fig.\ \ref{fig:para} also relates the model parameters to the leading-order behavior of the potential $\tilde{U}(\phi)$ for large $\phi$. This is interesting because it checks the boundedness of the potential. Although there is no obvious artifact in our results if the potential is unbounded it is useful to point out that this does occur for small values of $d$ and not too large values of $M_0$, see left panel of Fig.\ \ref{fig:para}. Unboundedness of the scalar potential after including the Dirac sea was also observed in quark-meson models \cite{Skokov:2010sf,Carignano:2014jla}; there, however, affecting the entire parameter space due to the different form of the tree-level potential.

The right panel of Fig.\ \ref{fig:phase_diagram} shows the case of a physical pion mass. Let us first compare the pale black curves (chiral limit, taken from the left panel) with the bold black curve (physical pion mass). First of all, we see that the second-order chiral phase transition line between the isotropic NM and $\chi$S phases disappears as the pion mass is switched on. This indicates that there is no strict distinction between nuclear matter and the chirally restored phase and a crossover is realized, as already discussed in Sec.\ \ref{sec:isotropic}. The CDW region is bounded by a first-order transition and it has retreated significantly compared to the second-order lines. This is in accordance with the observation of Fig.\ \ref{fig:specific_case}, where we have seen that the explicit chiral symmetry breaking tends to disfavor the CDW. With the most likely empirical range $M_0=(0.7-0.8)m_N$ in mind, we see from the black curve that the CDW may just about be realized, if $M_0$ is on the upper end of this range. Again, it is useful to consult Fig.\ \ref{fig:para} to get an idea of the corresponding values of $L$ and $m_\sigma$. For instance, for $M_0=0.81 m_N$ (the case used in Sec.\ \ref{sec:res_chiral_limit}) we have $L\simeq 54\, {\rm MeV}$ and $m_\sigma\simeq 1.1\, {\rm GeV}$, which is in tension with the empirically expected value of the sigma mass if the sigma is identified with the $f_0(500)$. The curves for different vector meson couplings (red, green blue) show that the CDW becomes more relevant for larger $d$, as already anticipated from the chiral limit in the left panel. Larger values of $d$ correspond to smaller $L$, well within experimental boundaries (perhaps even closer to the real-world value, judging from the distribution of experimental results), but also to larger values of the sigma mass. 

Additionally, the plot indicates that the CDW can only appear at  large baryon densities (markers on the CDW transition curves). The lowest possible densities are about $n_B\sim 6 n_0$, and these are only realized for large,  perhaps unrealistically large,  $M_0$. (Recall that $M_0$ has an upper bound (\ref{M0limit}), slightly above the scale shown here; as this bound is approached, $g_\omega$ goes to zero, which decreases the sensitivity of the results on $\omega$ and thus on $d$.) More realistic values of $M_0$ require increased values of $d$, leading to even higher baryon densities for the CDW onset. We have checked that large $d$ generally induce high densities at moderate values of the chemical potential. These large number susceptibilities suggest that the parameter regions where our model predicts a CDW produce soft equations of state. Therefore, it is possible that in these parameter regions the model predicts maximum masses of neutron stars incompatible with astrophysical observations. This remains to be verified by computing the mass-radius curve  under neutron star conditions, going beyond the isospin-symmetric scenario considered here.

\subsection{Comparison with different approaches to the Dirac sea}\label{sec:compare}

\begin{figure}
\begin{center}
\hbox{\includegraphics[width=0.5\textwidth]{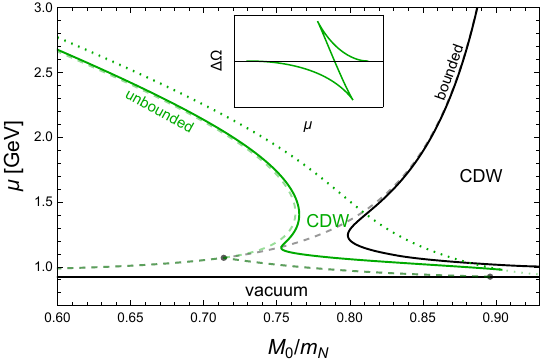}
\includegraphics[width=0.5\textwidth]{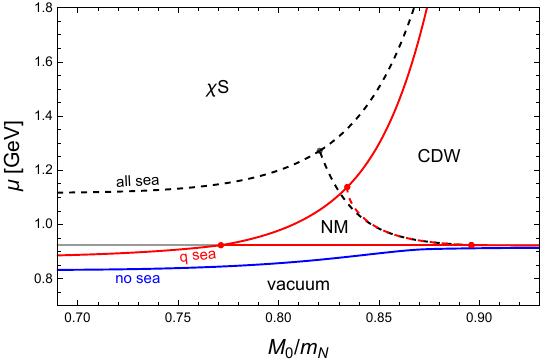}}
\caption{Comparison of our results to other approaches regarding the nucleonic vacuum contribution, in the plane of the model parameter $M_0$ and the chemical potential $\mu$. In both panels $K=250\, {\rm MeV}$, while $d=10^4$ (left) $d=50$ (right). {\it Left panel:} Results for the choice of the scale $\ell=m_N$, leading to an unbounded free energy (green), compared to our choice in Eq.\ (\ref{ellchoice}) (black, copied from the right panel of Fig.\ \ref{fig:phase_diagram}). Pale lines in both colors correspond to the chiral limit, bold lines to the physical pion mass. The inset shows the unphysical behavior in the free energy difference to the isotropic phase $\Delta\Omega$ for the unbounded case. The dotted lines (pale and bold lines essentially on top of each other) correspond to the point where $\Delta\Omega$ is minimal. {\it Right panel:} Effect of the Dirac sea in the chiral limit. The full calculation (``all sea'') is shown in black.  The red curves are obtained by dropping the $q=0$ contributions to the Dirac sea (``q sea''). If the entire Dirac sea contribution is dropped (``no sea'') the only transition is from the vacuum directly to the CDW phase (blue line), and the CDW persists for arbitrarily large $\mu$. In this panel, the two cases that include the Dirac sea are computed with the renormalization scale from Eq.\ (\ref{ellchoice}) (this choice is only relevant for the CDW-$\chi$S transitions, where $q>0$). In both panels, solid (dashed) lines are first (second) order phase transitions.}
\label{fig:compare}
\end{center}
\end{figure} 

Finally, let us compare our findings with two different treatments of the Dirac sea: firstly, in the left panel of Fig.\ \ref{fig:compare}, the use of a different renormalization scale and, secondly, in the right panel, neglecting the Dirac sea partially or altogether. Both comparisons are useful to relate our work to previous studies and are relevant to future improvements in different models.

In Refs.\ \cite{Carignano:2011gr,Carignano:2014jla} it was pointed out that in the NJL model and in particular the renormalizable quark-meson model, there is a curious 
behavior at large chemical potentials if the Dirac sea is taken into account: Depending on the parameters of the model, a re-entrance to the CDW phase can occur and this CDW ``island'' ends at an unphysical boundary. Here, ``unphysical'' means that the CDW solution turns around and continues back to smaller chemical potentials at a point where it is the favored phase. This predicts an unphysical jump in the free energy from the CDW to the chirally restored phase, see inset in the left panel of Fig.\ \ref{fig:compare}. This panel demonstrates that we find exactly the same behavior if the renormalization scale is chosen to be $\ell = m_N$ rather than choosing the $q$-dependent scale (\ref{ellchoice}): For $\ell=m_N$ (green curves) there are various different scenarios, depending on the value of the model parameter $M_0$, but in each case the CDW phase has an upper unphysical boundary (dotted line) as just described. (At very large $M_0$ there is no CDW region at all for a physical pion mass and $\ell=m_N$.) 

It is not surprising that the renormalization scale plays a crucial role here: Our choice, as argued at the end of Sec.\ \ref{sec:free}, was motivated by avoiding unboundedness of the free energy in the $q$-direction. This unboundedness, in turn, was identified as a problem in Refs.\ \cite{Broniowski:1990gb,Carignano:2014jla} (but not fixed by a suitable renormalization scheme), and it was realized in Ref.\ \cite{Carignano:2014jla} that the unboundedness contributes to the unphysical behavior, here shown by the green curves. Due to the close similarity between the quark-meson model and our nucleon-meson model, our results suggest  that the same renormalization procedure can remove the artifact in the quark-meson model. 

As mentioned already in Sec.\ \ref{sec:free}, the scale $\ell$ can be chosen differently while still maintaining a bounded free energy. For instance, we can generalize $\ell=\sqrt{m_N^2 + (2cq)^2}$, with a numerical factor $c$, which has a lower limit $c=1/\exp(1+2\pi^2/g_\sigma^2)\simeq 0.30$
if boundedness is required and, as $c\to 0$, connects our results continuously to the  case $\ell=m_N$. In this paper, we do, for definiteness, not further explore the dependence  of our results with respect to such variations. It should therefore be kept in mind that the phase boundaries of the CDW phase acquire some uncertainty in our scheme, which becomes larger for larger values of $q$ (corresponding to large $\mu$), and which may be alleviated by more elaborate approximations beyond the mean-field approach.    

Turning to the second aspect of this section, we now compare our results to the no-sea approximation, which was used in all previous works on the CDW in nucleonic  models. This comparison is done in the right panel of  Fig.\ \ref{fig:compare}. For convenience, we perform this comparison in the chiral limit because in this case all transitions are clearly visible as phase transition lines, crossovers being excluded. We distinguish two different  approximations. The red curves are obtained by dropping only the $q=0$ sea contribution. This amounts to setting $\tilde{U}=U$, i.e., dropping the difference between $\tilde{U}$ and $U$ in Eq.\ (\ref{Utilde}), but keeping all terms in $\Delta\tilde{U}$ (\ref{DelUtilde}). This approximation, labeled by ``q sea'' in Fig.\ \ref{fig:compare},  is reminiscent of the one used in Ref.\ \cite{Haber:2014ula}, where rotational symmetry is broken by an external magnetic field $B$ instead of the CDW and it was argued that the $B$-dependent vacuum contribution contains all important physics, while the $B=0$ vacuum contribution can be ignored without changing the results qualitatively. We have already seen that if we are interested in the chiral phase transition (which was not relevant in Ref.\ \cite{Haber:2014ula}), already the isotropic calculation is affected by the Dirac sea, turning the first-order chiral transition into a crossover. 

Nevertheless, in Fig.\ \ref{fig:compare} we see that the q-sea approximation reproduces many of the features of the full result. In contrast, if the entire sea contribution is omitted, $\tilde{U}=U$ and $\Delta\tilde{U}=\Delta U$, the result changes dramatically (blue curve). In that case, the behavior is qualitatively the same for all values of $M_0$: there is a first-order transition from the vacuum to the CDW, and the CDW persists for all values of the chemical potential, i.e., all isotropic phases with nonzero baryon number are gone. If we repeat the calculation for a physical pion mass (not shown in the plot) we find a parameter region, $0.87\lesssim M_0/m_N\lesssim0.93$, where the blue line moves above the baryon onset at $\mu=\mu_0$, opening up a pocket of isotropic nuclear matter. This is the scenario (vacuum $\to$ nuclear matter $\to$ CDW) found in Ref.\ \cite{Heinz:2013hza} in a similar model within the no-sea approximation, investigating only one specific parameter set. With our more global view of the parameter space we conclude that the no-sea approximation vastly overestimates the importance of the CDW, while the q-sea approximation is much closer to the full result, which takes into account the entire nucleonic vacuum contribution.

\section{Summary and Outlook}\label{sec:summary}

We have employed a nucleon-meson model to improve earlier studies on the possibility of an anisotropic chiral condensate in dense, isospin-symmetric  nuclear matter. The model is based on nucleonic degrees of freedom which interact via meson exchange. Importantly, the fermion masses are generated dynamically such that the model can be used to study the chiral phase transition. In our ansatz for the anisotropic chiral condensate we have restricted ourselves to the CDW, which does not break translational invariance. We have worked at zero temperature and in the mean-field approximation.

An important part of our study has been the nucleonic vacuum contribution. We have argued that this contribution is already crucial in the isotropic scenario: it turns the first-order chiral transition into a crossover. As a consequence, our main results concern the question whether the CDW disrupts the smooth transition from nuclear matter to approximately chirally restored matter. We have found that this is indeed possible and have discussed the dependence of the CDW region on the model parameters. By studying the chiral limit as well as the case of a physical pion mass, we have shown that the CDW tends to be disfavored by explicit chiral symmetry breaking. Independent of the choice of the parameters, we have found that within our model the CDW can only appear at large baryon densities, $n_B\gtrsim 6 n_0$. It is realized somewhere at the edges of and beyond the parameter regime empirically allowed by nuclear saturation properties.

On a more theoretical note, we have discussed a renormalization scheme, and in particular a certain choice of the renormalization scale, which fixes a problem pointed out in similar models based on quark degrees of freedom. Within our scheme, there is no re-entrance and/or unphysical behavior of the CDW at ultra-high densities and it would be interesting to apply our scheme -- possibly in modified or further improved form -- also to different phenomenological or effective models that describe the CDW or related non-uniform phases. 

Our work also opens up other directions for future work. Most straightforwardly, one can think of improvements and extensions of our calculation, for instance including nonzero-temperature effects and/or mesonic vacuum fluctuations. Somewhat more substantial extensions would be an improved ansatz for the anisotropic chiral condensate, possibly a comparison of different inhomogeneous structures, or taking into account Cooper pairing. One could also repeat our calculation in the extended linear sigma model of Ref.\ \cite{Heinz:2013hza} -- where the CDW was studied previously, but without the effect of the Dirac sea -- or with the inclusion of strangeness along the lines of Ref.\ \cite{Fraga:2022yls} or \cite{Fraga:2023wtd}. Applications of our results to real-world systems concern the interior of neutron stars. It would therefore be interesting to generalize our calculation to isospin-asymmetric matter and to impose the conditions of electric neutrality and equilibrium with respect to the electroweak interaction. In this context one may study the competition or possible coexistence of the CDW with quark-hadron mixed phases, which become conceivable due to the presence of a second chemical potential associated with electric charge. One can then proceed by predicting the size of the layer inside the star where a CDW or a more complicated spatial structure can be expected and relate this finding to astrophysical data, perhaps via transport properties such as the neutrino emissivity.

\section{Acknowledgments}
We thank Mark Alford, Michael Buballa, Stefano Carignano, Eduardo Fraga, Rodrigo da Mata, and Dirk Rischke for valuable comments and discussions, and  Orestis Papadopoulos for pointing out typos in two equations in a previous version of the manuscript.

\appendix

\section{Computing the Dirac sea contribution}
\label{sec:diracsea}

\subsection{Regularization}
\label{sec:regularization}

In order to regularize the divergent part of the baryonic pressure $P_{\rm vac}$ (\ref{D0def}) we employ proper time regularization. First, we use 
\begin{equation}
\frac{1}{x^a} = \frac{1}{\Gamma(a)}\int_0^\infty d\tau\, \tau^{a-1}e^{-\tau x}
\end{equation}
to rewrite $E_k^s$ from Eq.~\eqref{eq:dispersion}, setting 
$a=-1/2$, $x= (\sqrt{k_\ell^2+M^2}+s q)^2+k_\perp^2$. We can then perform the $k_\perp$ integration to obtain 
\begin{align}\label{D01}
P_{\rm vac} = -\frac{1}{4\pi^{5/2}}\int_0^\infty\frac{d\tau}{\tau^{5/2}}\int_0^\infty dk_\ell\, e^{-\tau(k_\ell^2+M^2+q^2)} 
\cosh 2q\tau\sqrt{k_\ell^2+M^2} \, .
\end{align}
Next, after inserting the series expansion
\begin{equation} \label{cosh}
\cosh x = \sum_{n=0}^\infty \frac{x^{2n}}{(2n)!} \, , 
\end{equation}
we can perform the $k_\ell$ integral to obtain 
\bea
P_{\rm vac} = \sum_{n=0}^\infty\int_{0}^\infty \frac{\tau^{2n}d\tau}{\tau^{5/2}}{\cal P}_n \, ,
\eea
where 
\bea
{\cal P}_n \equiv -\frac{M}{8\pi^2}\frac{(2qM)^{2n}}{(2n)!}  e^{-\tau(M^2+q^2)}  \Psi\left(\frac{1}{2},\frac{3}{2}+n,\tau M^2\right) \, ,
\eea
with the confluent hypergeometric function of the second kind $\Psi(a,b,z)$.

For small $\tau$ we have 
\begin{equation}
\frac{\tau^{2n}}{\tau^{5/2}}  \Psi\left(\frac{1}{2},\frac{3}{2}+n,\tau M^2\right) \propto \tau^{n-3} \, .
\end{equation}
Therefore, the $\tau$ integral is finite for $n\ge 3$.
For $n=0,1,2$ we replace the lower boundary by a cutoff $1/\Lambda^2$ to compute
\bea \label{D0inf}
 \sum_{n=0}^2\int_{1/\Lambda^2}^\infty \frac{\tau^{2n}d\tau}{\tau^{5/2}}{\cal P}_n
&=&  -\frac{\Lambda^4}{16\pi^2}+\frac{\Lambda^2M^2}{8\pi^2}+ \frac{M^4}{16\pi^2}\left(\gamma-\frac{3}{2}+\ln\frac{M^2+q^2}{\Lambda^2}\right)+\frac{q^2M^2}{4\pi^2}\left(\gamma+\ln\frac{M^2+q^2}{\Lambda^2}\right) \non[2ex]
&&+\frac{q^4}{96\pi^2}\frac{3-8y^2-25y^4-6y^6}{(1+y^2)^2}+{\cal O}\left(\frac{1}{\Lambda^2}\right)  \, , 
\eea
where $\gamma\simeq 0.577$ is the Euler-Mascheroni constant and we have used the abbreviation $y$ as defined in Eq.\ (\ref{ydef}). 

For the terms $n\ge 3$ it is easier to go back to the original expression (\ref{D01}), insert the series (\ref{cosh}), and then first perform the $\tau$ integral. With the new integration variables $k_\ell'=k_\ell/q$, $\tau'=q^2\tau$, abbreviating 
\be
\kappa^2 \equiv k_\ell'^2+y^2 \, , 
\ee
and dropping the primes again for convenience,
we compute  
\bea \label{D0fin}
&&-\frac{q^4}{4\pi^{5/2}}\int_0^\infty dk_\ell\sum_{n=3}^\infty \frac{(2\kappa)^{2n}}{(2n)!} \int_{0}^\infty\frac{\tau^{2n} d\tau}{\tau^{5/2}}\,e^{-\tau(\kappa^2+1)} 
=-\frac{q^4}{4\pi^{5/2}}\int_0^\infty dk_\ell\sum_{n=3}^\infty \frac{\Gamma(2n-3/2)}{(2n)!\sqrt{\pi}} \frac{(2\kappa)^{2n}}{(\kappa^2+1)^{2n-3/2}} \non[2ex]
&& = -\frac{q^4}{6\pi^{2}}\int_0^\infty dk_\ell \left[\left(1+\kappa\right)^3+\left|1-\kappa\right|^3 -\frac{3\kappa^4+12\kappa^2
(1+\kappa^2)^2+8(1+\kappa^2)^4}{4(1+\kappa^2)^{5/2}}\right] \hspace{1cm}\non[2ex]
&&= -\left(\frac{M^4}{16\pi^2}+\frac{q^2M^2}{4\pi^2}\right)\ln\frac{M^2+q^2}{M^2}+\frac{q^4}{96\pi^2}\frac{5+24y^2+33y^4+6y^6}{(1+y^2)^2}\non[2ex]
&&+\frac{q^4\Theta(1-y)}{4\pi^2}\left[-\sqrt{1-y^2}\frac{2+13y^2}{6}+2y^2\left(1+\frac{y^2}{4}\right)\ln\frac{1+\sqrt{1-y^2}}{y} \right] \, .
\eea
Adding the results (\ref{D0inf}) and (\ref{D0fin}), we obtain the compact expression 
\begin{equation}\label{eq:D0regularized}
P_{\rm vac} = -\frac{\Lambda^4}{16\pi^2}+\frac{\Lambda^2M^2}{8\pi^2}+\frac{M^4}{16\pi^2}\left(\gamma-\frac{3}{2}+\ln\frac{M^2}{\Lambda^2}\right)+\frac{q^2M^2}{4\pi^2}\left(\gamma+\ln\frac{M^2}{\Lambda^2}\right)+\frac{q^4}{4\pi^2}F(y) +{\cal O}\left(\frac{1}{\Lambda^2}\right) \, , 
\end{equation}
with $F(y)$ defined in Eq.\ (\ref{Fdef}).

\subsection{Renormalization}
\label{sec:renormalization}

Removing the divergences in Eq.~\eqref{eq:D0regularized} requires renormalization. To this end, we first introduce a renormalization scale $\ell$ and drop the terms of order $1/\Lambda^2$ and higher to rewrite Eq.\ (\ref{eq:D0regularized}) as 
\bea \label{D02}
-2P_{\rm vac} &=& \frac{\Lambda^4}{8\pi^2}-\frac{\Lambda^2M^2}{4\pi^2}- \left(\frac{M^4}{8\pi^2}+\frac{q^2M^2}{2\pi^2}\right)\left(\gamma-\frac{3}{2}+\ln\frac{\ell^2}{\Lambda^2}\right) - \frac{M^4}{8\pi^2}\ln\frac{M^2}{\ell^2}-\frac{q^2M^2}{2\pi^2}\left(\ln\frac{M^2}{\ell^2}+\frac{3}{2}\right)-\frac{q^4}{2\pi^2}F(y) \, . \non 
\eea
We have also reinstated the isospin degeneracy factor 2 and a minus sign to obtain the total vacuum contribution from neutrons and protons to the free energy, cf.\ Eq.\ (\ref{Ombar}).  

Next, we interpret the following fields and parameters in the Lagrangian as bare quantities, related to the corresponding renormalized quantities via 
\begin{equation}\label{eq:rescaling}
\phi = Z^{1/2}\phi_r \, , \qquad f_\pi = Z^{1/2}f_{\pi, r} \, , \qquad g_\sigma = \frac{g_{\sigma,r}}{Z^{1/2}} \, , \qquad a_n = \frac{a_{n,r}+f_{\pi,r}^{4-2n}\,\delta a_n}{Z^{n}}  \, , \qquad \epsilon = \frac{\epsilon_r}{Z^{1/2}} \, , 
\end{equation}
where we have introduced the dimensionless field rescaling factor $Z$ and the dimensionless counterterms $\delta a_n$. The rescaling of $\phi $ follows from rescaling $\sigma = Z^{1/2} \sigma_r,\; \pi_a = Z^{1/2}\pi_{a,r}$ in the original Lagrangian. 
The remaining fields and parameters of the Lagrangian are assumed to be already in their renormalized form. Therefore, the only terms in the mean-field Lagrangian (\ref{LMF}) affected by the renormalization (\ref{eq:rescaling}) are 
\bea \label{UDelUr}
U + \Delta U&=&  \sum_{n=1}^4 \frac{a_{n,r}+f_{\pi,r}^{4-2n}\,\delta a_n}{n!}\frac{(\phi_r^2-f_{\pi,r}^2)^n}{2^n} -\epsilon_r(\phi_r-f_{\pi,r}) + 2Z\phi_r^2q^2+(1-\delta_{0q})\epsilon_r\phi_r  \non[2ex]
&=&(U+\Delta U)_r+f_{\pi,r}^4\left[\left(-\frac{\delta a_1 }{2}+\frac{\delta a_2  }{8}-\frac{\delta a_3}{48}+\frac{\delta a_4 }{384} \right)+\left(\frac{\delta a_1}{2}-\frac{\delta a_2}{4}+\frac{\delta a_3 }{16}-\frac{\delta a_4 }{96} \right)\varphi^2\right.\non[2ex]
&&\left.+\left(\frac{\delta a_2}{8}-\frac{\delta a_3 }{16}+\frac{\delta a_4}{64} \right)\varphi^4+\left(\frac{\delta a_3}{48}-\frac{\delta a_4 }{96}\right)\varphi^6+
\frac{\delta a_4}{348 }\varphi^8\right]+2 (Z-1)\phi_r^2 q^2  \, , 
\eea
where $(U+\Delta U)_r$ is given by $U$ and $\Delta U$ from Eq.\ (\ref{UDelU}) with $\phi,f_\pi,a_n,\epsilon$ replaced by their renormalized versions, and where $\varphi$ is defined in Eq.\ (\ref{varphidef}).

We observe from Eq.\ (\ref{D02}) that we need to cancel divergent terms in $P_{\rm vac}$ proportional to $M^2$, $M^4$, and $q^2M^2$.  Since $M$ and $q$ are dynamical quantities that depend on the medium, this cancelation has to be done order by order with the help of the counterterms in Eq.\ (\ref{UDelUr}). To make the cancelation explicit we divide the counterterms and the field rescaling into divergent and finite parts,
\begin{equation}
\delta a_n = \delta a_n^\Lambda + \delta a_n^\text{f} \, , \qquad  Z = Z^\Lambda + Z^\text{f} \, .
\end{equation}
The divergent terms proportional to $M^2$ and $M^4$ are then canceled (and no new divergences introduced) by the choice 
\begin{equation}\label{eq:danL}
\delta a_1^\Lambda = \frac{g_{\sigma,r}^4}{2\pi^2}\left(\frac{\Lambda^2}{m_N^2} +  \ln\frac{\ell^2}{\Lambda^2} +\gamma-\frac{3}{2} \right) \, , \qquad 
\delta a_2^\Lambda =\frac{g_{\sigma,r}^4}{\pi^2} \left(\ln \frac{\ell^2}{\Lambda^2}+\gamma-\frac{3}{2}\right) \, , \qquad \delta a_3^\Lambda = \delta a_4^\Lambda =0 \, , 
\end{equation}
while the divergent term proportional to $q^2M^2$ is canceled by 
\begin{equation}\label{eq:zL}
Z^\Lambda = \frac{g_{\sigma,r}^2}{4\pi^2}\left(\ln \frac{\ell^2}{\Lambda^2}+\gamma-\frac{3}{2}\right) \, .
\end{equation}
Besides the divergent terms, the vacuum contribution (\ref{D02}) also contains finite logarithmic terms, with prefactors $M^4$ and $q^2M^2$. Let us start with the logarithmic term with prefactor $M^4$. We combine this contribution with the finite part of the counterterms $\delta a_n^{\rm f}$. While for the identification of the divergent parts of the counterterms we applied an expansion in $\varphi$ (\ref{UDelUr}), we now expand about the vacuum, i.e., in $\varphi^2-1$, to write 
\be \label{anf1}
\sum_{n=1}^4 \frac{a_{n,r}+f_{\pi,r}^{4-2n}\,\delta a_n^{\rm f}}{n!}\frac{(\phi_r^2-f_{\pi,r}^2)^n}{2^n}- \frac{M^4}{8\pi^2}\ln\frac{M^2}{\ell^2} = -\frac{m_N^4}{8\pi^2}\ln\frac{m_N^2}{\ell^2} + \sum_{n=1}^4 \frac{A_n}{n!}\frac{(\varphi^2-1)^n}{2^n} + \frac{m_N^4}{4\pi^2}\sum_{n=5}^\infty\frac{(-1)^n(\varphi^2-1)^n}{n(n-1)(n-2)} \, , 
\ee
where
\bea\allowdisplaybreaks
A_1 &\equiv& f_{\pi,r}^{2}a_{1,r} +f_{\pi,r}^4\left[\delta a_1^{\rm f}-\frac{g_{\sigma,r}^4}{4\pi^2}\left(1+2\ln\frac{m_N^2}{\ell^2}\right)\right] \, , \qquad 
A_2 \equiv f_{\pi,r}^{4}a_{2,r} +f_{\pi,r}^4\left[\delta a_2^{\rm f}-\frac{g_{\sigma,r}^4}{2\pi^2}\left(3+2\ln\frac{m_N^2}{\ell^2}\right)\right] \, , \non[2ex]
A_3 &\equiv& f_{\pi,r}^{6}a_{3,r} +f_{\pi,r}^4\left(\delta a_3^{\rm f}-\frac{2g_{\sigma,r}^4}{\pi^2}\right) \, , \qquad 
A_4 \equiv f_{\pi,r}^{8}a_{4,r} +f_{\pi,r}^4\left(\delta a_4^{\rm f}+\frac{4g_{\sigma,r}^4}{\pi^2}\right) \, .
\eea
The new coefficients $A_n$ entirely encode the form of the scalar potential and they will be fixed to physical properties of the vacuum and saturated nuclear matter. As a consequence, the choice of the renormalization scale and the finite counterterms is irrelevant here; for any particular choice of $\ell$ and $\delta a_n^{\rm f}$ the coefficients $a_{n,r}$ can be readjusted to reproduce the desired values for $A_n$. This  implies that the form of the original mesonic potential, which contains terms $(\varphi^2-1)^n$ for $n=1,2,3,4$, is not altered by the renormalization scheme, although the coefficients of these terms will assume different values due to the Dirac sea. The reason is the presence of the higher-order terms $(\varphi^2-1)^n$ for $n\ge 5$, given by the last term in Eq.\ (\ref{anf1}). They do not depend on any free parameters and cannot be eliminated by any choice of the renormalization scale or the counterterms. We can rewrite this infinite sum in the closed form 
\be
\frac{m_N^4}{4\pi^2}\sum_{n=5}^\infty\frac{(-1)^n(\varphi^2-1)^n}{n(n-1)(n-2)} = \frac{m_N^4}{96\pi^2}(1-8\varphi^2-12\varphi^4\ln\varphi^2 +8\varphi^6 -\varphi^8) \, .
\ee
Next, we consider the logarithmic term with a $q$-dependent prefactor in Eq.\ (\ref{D02}). Combining this term with the finite part of the field rescaling from Eq.\ (\ref{UDelUr}), we write 
\bea \label{Zf1}
2(Z^{\rm f}-1)\phi_r^2q^2-\frac{q^2M^2}{2\pi^2}\left(\ln\frac{M^2}{\ell^2}+\frac{3}{2}\right)  &=& -\frac{q^2M^2}{2\pi^2}\ln\frac{M^2}{\ell^2} \, , 
\eea
where we have set 
\be
Z^{\rm f} = 1+\frac{3g_{\sigma,r}^2}{8\pi^2} \, .
\ee
This choice leaves a renormalization scale dependence, in contrast to the case of the $q$-independent contribution. As we discuss in the main part of the paper, this renormalization scale dependence gives us an important freedom to eliminate unphysical properties of our effective potential. 

Putting everything together, we can write 
\be \label{D0UDelU}
-2P_{\rm vac} + U + \Delta U = \frac{\Lambda^4}{8\pi^2}-\frac{\Lambda^2m_N^2}{4\pi^2} -\frac{m_N^4}{8\pi^2}\left(\ln\frac{m_N^2}{\Lambda^2}-\frac{3}{2}+\gamma\right) + \tilde{U} + \Delta\tilde{U} \, , 
\ee
where we have absorbed the effects from the nucleonic Dirac sea into a new effective potential, given by 
\begin{subequations}
\bea
\tilde{U} &=& \sum_{n=1}^4 \frac{A_n}{n!}\frac{(\varphi^2-1)^n}{2^n}-\epsilon_r(\phi_r-f_{\pi,r})+\frac{m_N^2}{96\pi^2}(1-8\varphi^2-12\varphi^4\ln\varphi^2 +8\varphi^6 -\varphi^8) \, , \\[2ex]
\Delta\tilde{U} &=& 2\phi_r^2q^2\left(1-\frac{g_{\sigma,r}^2}{4\pi^2}\ln\frac{M^2}{\ell^2}\right) -\frac{q^4}{2\pi^2}F(y)+(1-\delta_{0q})\epsilon_r\phi_r \, .
\eea
\end{subequations}
As for the original potential, we have separated the $q$-dependent part $\Delta \tilde{U}$ such that the potential reduces to $\tilde{U}$ for $q=0$.  
Dropping the irrelevant (divergent, but constant) terms in Eq.\ (\ref{D0UDelU}), denoting the renormalized quantities for simplicity without the subscript $r$ and renaming $A_n/f_{\pi}^{2n} \to a_{n}$,  we arrive at the result (\ref{Omren}) given in the main text.

\section{Matter contributions to densities and axial current}\label{ap:functions}

In this appendix we present the explicit expressions for the matter contributions to the stationarity equations (\ref{stat}). 
The baryon density from a single nucleonic degree of freedom is  
\bea
\frac{\partial P_{\rm mat}}{\partial \mu} &=&\frac{1}{2\pi^2}\sum_{s=\pm} \int_0^\infty dk_\ell\int_0^\infty dk_\perp k_\perp \Theta(\mu_*-E_k^s) \non[2ex]
&=& -\frac{\Theta (\mu_* -q -M)}{4\pi^2}\left\{M^2q\ln \frac{\mu_*-q+k_-}{M}+\frac{k_-}{3} [ 2(M^2-\mu_*^2) +q(q+\mu_*)]\right\} \non[2ex]
&& +\frac{\Theta (\mu_* +q -M)}{4\pi^2}\left\{M^2q\ln \frac{\mu_*+q+k_+}{M}-\frac{k_+}{3} [ 2(M^2-\mu_*^2) +q(q-\mu_*)]\right\} \non[2ex]
&& -\frac{\Theta (q -\mu_* -M)}{4\pi^2}\left\{M^2q\ln \frac{q-\mu_*+k_-}{M}-\frac{k_-}{3} [ 2(M^2-\mu_*^2) +q(q+\mu_*)]\right\} \, , 
\eea
with $k_\pm$ from Eq.\ (\ref{kpm}). To obtain the baryon density $n_B$ in the stationarity equation (\ref{eq:omega}) the result has to be multiplied by 2 due to the (degenerate) contributions from neutrons and protons.
One easily checks that one obtains the expected limits 
\bea
\frac{\partial P_{\rm mat}}{\partial \mu} = \left\{
\begin{array}{cc} \displaystyle{\frac{\Theta(\mu_*-M)k_F^3}{3\pi^2}} & \;\;\mbox{for}\;\; q=0 \\[2ex]
\displaystyle{\frac{\mu_*^3}{3\pi^2}} & \;\;\mbox{for}\;\; M=0 \end{array}
\right.
\eea
In particular, the density does not depend on $q$ for zero fermion mass $M=0$. 

The scalar density is given by 
\bea \allowdisplaybreaks
 -\frac{\partial P_{\rm mat}}{\partial M}&=& \frac{1}{2\pi^2}\sum_{s=\pm} \int_0^\infty dk_\ell\int_0^\infty dk_\perp k_\perp \frac{M}{E_k^s}\left(1+\frac{sq}{\sqrt{k_\ell^2+M^2}}\right)\Theta(\mu-E_k^s)\non[2ex]
&=&-\frac{\Theta(\mu_*-q-M)M}{4\pi^2}\left\{[M^2+2q(q-\mu_*)]\ln\frac{\mu_*-q+k_-}{M}-(\mu_*-3q)k_-\right\} \non[2ex]
&&-\frac{\Theta(\mu_*+q-M)M}{4\pi^2}\left\{[M^2+2q(q+\mu_*)]\ln\frac{\mu_*+q+k_+}{M}-(\mu_*+3q)k_+\right\} \non[2ex]
&&-\frac{\Theta(q-\mu_*-M)M}{4\pi^2}\left\{[M^2+2q(q-\mu_*)]\ln\frac{q-\mu_*+k_-}{M}+(\mu_*-3q)k_-\right\} \non[2ex]
&&+\frac{\Theta(q-M)M}{2\pi^2}\left[ (M^2+2q^2)\ln \frac{q+\sqrt{q^2-M^2}}{M}-3q\sqrt{q^2-M^2}\right] \, .
\eea
In this case, we recover the well-known expression for $q=0$, 
\bea
-\frac{\partial P_{\rm mat}}{\partial M}=\frac{\Theta(\mu_*-M)M}{2\pi^2}\left(\mu_* k_F-M^2\ln\frac{\mu_*+k_F}{M}\right) \, , 
\eea
while for small $M$ we find the expansion 
\bea
-\frac{\partial P_{\rm mat}}{\partial M} = \frac{M}{2\pi^2}\left(\mu_*^2-\mu_* q\ln\left|\frac{\mu_*+q}{\mu_*-q}\right|-q^2\ln\left|\frac{\mu_*^2}{q^2}-1\right|\right)+{\cal O}(M^3) \, , 
\eea
which confirms that Eq.\ (\ref{eq:sigma}) is solved by $M=0$ in the chiral limit $\epsilon=0$. 

Finally, the axial current from a single nucleonic degree of freedom is
\bea \label{dPdq}
 -\frac{\partial P_{\rm mat}}{\partial q}&=&\frac{1}{2\pi^2}\sum_{s=\pm} s\int_0^\infty dk_\ell\int_0^\infty dk_\perp k_\perp \frac{\sqrt{k_\ell^2+M^2}+sq}{E_k^s}\Theta(\mu-E_k^s)\non[2ex]
&=&\frac{\Theta(\mu_*-q-M)}{4\pi^2}\left[M^2(\mu_*-2q)\ln \frac{\mu_*-q+k_-}{M}-\frac{k_-}{3}(4M^2-\mu_*^2-\mu_* q +2q^2)\right]\non[2ex]
&&-\frac{\Theta(\mu_*+q-M)}{4\pi^2}\left[M^2(\mu_*+2q)\ln \frac{\mu_*+q+k_+}{M}-\frac{k_+}{3}(4M^2-\mu_*^2+\mu_* q +2q^2)\right]\non[2ex]
&&+\frac{\Theta(q-\mu_* -M)}{4\pi^2}\left[M^2(\mu_*-2q)\ln \frac{q-\mu_*+k_-}{M}+\frac{k_-}{3}(4M^2-\mu_*^2-\mu_* q +2q^2)\right]\non[2ex]
&&+\frac{\Theta(q-M)}{\pi^2}\left[M^2 q\ln \frac{q+\sqrt{q^2 -M^2}}{M}-\frac{\sqrt{q^2-M^2}}{3}(2M^2+q^2)\right] \, .
\eea
The current is linear in $q$ for small $q$, 
\bea
-\frac{\partial P_{\rm mat}}{\partial q}
= -q\frac{\Theta(\mu_*-M)M^2}{\pi^2}\ln\frac{\mu_*+k_F}{M} + {\cal O}(q^2) \, , 
\eea
while it is quadratic in $M$ for small $M$,
\bea
-\frac{\partial P_{\rm mat}}{\partial q}
= -\frac{M^2}{2\pi^2}\left(\frac{\mu_*}{2}\ln\left|\frac{\mu_*+q}{\mu_*-q}\right|+q\ln\left|\frac{\mu_*^2}{q^2}-1\right|\right) +{\cal O}(M^4)  \, .
\eea

\section{Model parameters}\label{ap:para}

In this appendix we, firstly, present -- for completeness and replicability -- the model parameters used for the specific cases discussed in Secs.\ \ref{sec:isotropic} and \ref{sec:res_chiral_limit}, see Table \ref{table:para}. 

\begin{table}[h]
\begin{tabular}{||c | c | c | c |c || c | c |c|c||c||} 
 \hline
   $g_{\omega}$ 
   & $a_2$ 
   & $a_3 [{\rm MeV}^{-2}]$ 
   & $a_4 [{\rm MeV}^{-4}]$  & $d$ 
   & $\;$$M_0/m_N$$\;$ 
   & $\;$$L [{\rm MeV}]$$\;$ 
   & $\;$$m_\sigma [{\rm MeV}] $$\;$ 
   & $\;$$m_\pi [{\rm MeV}]$$\;$ 
   & $\;$sea$\;$ \\ [0.5ex] 
 \hline\hline
  $\;$ 7.574 $\;$ & $\;$59.94$\;$ & $\;$$-9.427\times 10^{-3}$$\;$ & $1.188\times 10^{-4}$ & 0 & 0.82& 87.3 & 708 & 0 & -- \\
  \hline
$\;$ 7.574 $\;$ & 57.75 & $-2.247\times 10^{-2}$ & $\;$$8.612\times 10^{-5}$$\;$ & 0 & 0.82& 87.3 & 708 & 0 & \checkmark \\
 \hline
 $\;$ 7.574 $\;$ & 57.66 & $-8.892\times 10^{-3}$ & $1.178\times 10^{-4}$ & 0 & 0.82& 87.3 & 707 & 139 & -- \\
 \hline
  $\;$ 7.574 $\;$ & 55.48 & $-2.193\times 10^{-2}$ & $8.512\times 10^{-5}$ & 0 & 0.82& 87.3 & 707 & 139 & \checkmark \\
 \hline\hline
 $\;$ 12.45 $\;$ & 130.8 & $0.4333$ & $7.850\times 10^{-4}$ & $\;$$10^4$$\;$ & 0.81& 53.8 & 1063 & 0 & \checkmark \\
 \hline
  $\;$ 12.45 $\;$ & 128.5 & $0.4338$ & $7.840\times 10^{-4}$ & $\;$$10^4$$\;$ & 0.81& 53.8 & 1063 & 139 & \checkmark \\
 \hline
\end{tabular}
\caption{Parameter sets together with resulting physical quantities used for the left panel of Fig.\ \ref{fig:isotropic}  (top four rows) and for Fig.\ \ref{fig:specific_case} (bottom two rows). In all cases, $K=250\, {\rm MeV}$, and the remaining vacuum and saturation properties not shown here are fixed to their physical values. To compute $L$ we always use a value for the symmetry energy of $S=32\, {\rm MeV}$. The last column indicates whether the Dirac sea is taken into account or not, which is relevant for the parameter fit.}
\label{table:para}
\end{table}

\begin{figure}[h]
\begin{center}
\hbox{\includegraphics[width=0.33\textwidth]{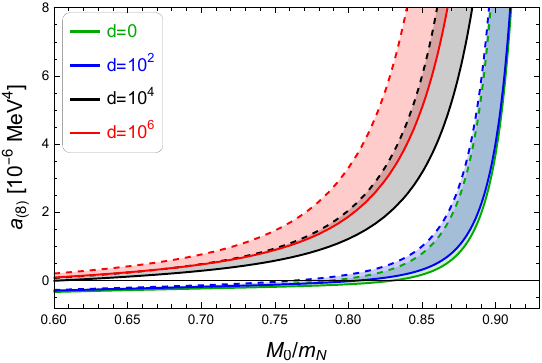}
\includegraphics[width=0.33\textwidth]{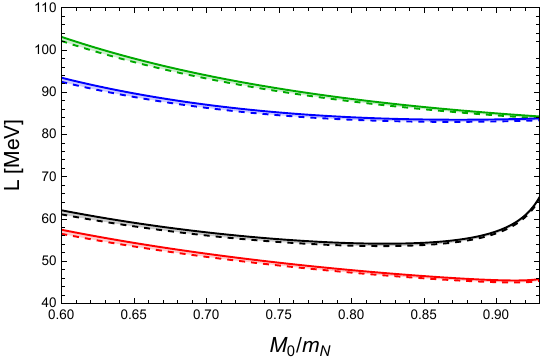}\includegraphics[width=0.33\textwidth]{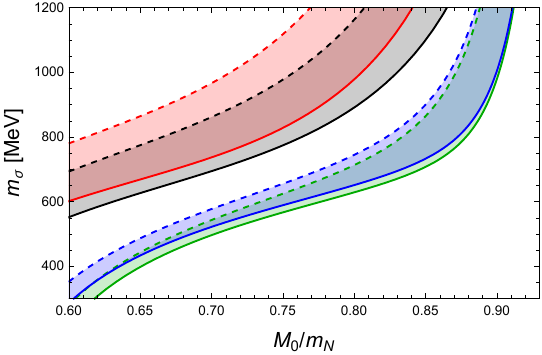}}
\caption{Leading-order coefficient of the effective potential $a_{(8)}$, slope parameter of the symmetry energy $L$, and sigma mass $m_\sigma$, as the effective nucleon mass at saturation $M_0$ is varied, with all other saturation properties kept fixed. In each panel,  the results for 4 different values of the vector meson self-coupling $d$ are shown, corresponding to the 4 values in the right panel of Fig.\ \ref{fig:phase_diagram}.  The bands indicate the range between $K=200\, {\rm MeV}$ (solid lines) and $K=300\, {\rm MeV}$ (dashed lines). All curves are calculated with the physical pion mass. The chiral limit gives slightly different curves but the differences would barely be visible on the scale of these plots. }
\label{fig:para}
\end{center}
\end{figure}

In our main results in Sec.\ \ref{sec:phase_structure}, the parameters are varied continuously. Therefore, secondly, we present the most relevant physical information about these continuous parameter sets in Fig.\ \ref{fig:para}.  This figure shows the slope parameter of the symmetry energy $L$ and the sigma mass $m_\sigma$ for different values of the vector meson self-coupling and the incompressibility as a function of the effective nucleon mass at saturation, computed from Eqs.\ (\ref{eq:L}) and (\ref{mpimsig}). Additionally, we show the coefficient of the leading-order term of the effective potential for large chiral condensates,
\be
\tilde{U}(\phi) = a_{(8)} \phi^8 + {\cal O}(\phi^6) \, , \qquad a_{(8)} \equiv \frac{1}{96}\left(\frac{a_4}{4}-\frac{g_\sigma^4}{\pi^2f_\pi^2}\right) \, .
\ee
The sign of $a_{(8)}$ indicates whether the potential is bounded from below for large $\phi$. The Dirac sea contribution is negative and thus tends to render the potential unbounded, which is indeed the case for small vales of $d$ and not too large values of $M_0$, as the figure demonstrates. 

\bibliography{references}

\end{document}